\documentclass[conference]{IEEEtran}
\IEEEoverridecommandlockouts

\usepackage{cite}
\usepackage{amsmath,amssymb,amsfonts}
\usepackage{graphicx}
\usepackage{siunitx}
\usepackage{textcomp}
\usepackage{xcolor}
\usepackage{subcaption}
\usepackage{algorithm}
\usepackage{algpseudocode}
\usepackage{multirow}
\usepackage[table]{xcolor}
\usepackage{booktabs}
\definecolor{CellFst}{HTML}{FF9999}
\definecolor{CellSnd}{HTML}{FFCC99}
\definecolor{CellTrd}{HTML}{FFFFFF}
\usepackage{bbm}
\usepackage{tikz}

\newcommand{\blackcircle}[1]{%
    \begin{tikzpicture}[baseline=(text.base)]
        \node[shape=circle, draw, fill=black, text=white, inner sep=1.5pt] (text) {#1};
    \end{tikzpicture}%
}

\def\BibTeX{{\rm B\kern-.05em{\sc i\kern-.025em b}\kern-.08em
    T\kern-.1667em\lower.7ex\hbox{E}\kern-.125emX}}
\begin{document}

\title{CoEdge-RAG: Optimizing Hierarchical Scheduling for Retrieval-Augmented LLMs in Collaborative Edge Computing

\thanks{* These authors contributed equally.}
\thanks{$\dagger$ Corresponding author.}
}



\author{
    \IEEEauthorblockN{\Large Guihang Hong$^{*}$, Tao Ouyang$^{*}$, Kongyange Zhao, Zhi Zhou, Xu Chen$^\dagger$}
    \IEEEauthorblockA{School of Computer Science and Engineering, Sun Yat-sen University, Guangzhou, China}
    \IEEEauthorblockA{\{honggh3, zhaokyg\}@mail2.sysu.edu.cn, \{ouyt33, zhouzhi9, chenxu35\}@mail.sysu.edu.cn}
}

    
    


\maketitle

\begin{abstract}
Motivated by the imperative for real-time responsiveness and data privacy preservation, large language models (LLMs) are increasingly deployed on resource-constrained edge devices to enable localized inference. To improve output quality, retrieval-augmented generation (RAG) is an efficient technique that  seamlessly integrates local data into LLMs. However, existing edge computing paradigms primarily focus on single-node optimization, neglecting opportunities to holistically exploit distributed data and heterogeneous resources through cross-node collaboration. To bridge this gap, we propose CoEdge-RAG, a hierarchical scheduling framework for retrieval-augmented LLMs in collaborative edge computing. In general, privacy constraints preclude accurate a priori acquisition of heterogeneous data distributions across edge nodes, directly impeding RAG performance optimization. Thus, we first design an online query identification mechanism using proximal policy optimization (PPO), which autonomously infers query semantics and establishes cross-domain knowledge associations in an online manner. Second, we devise a dynamic inter-node scheduling strategy that balances workloads across heterogeneous edge nodes by synergizing historical performance analytics with real-time resource thresholds. Third, we develop an intra-node scheduler based on online convex optimization, adaptively allocating query processing ratios and memory resources to optimize the latency-quality trade-off under fluctuating assigned loads. Comprehensive evaluations across diverse QA benchmarks demonstrate that our proposed method significantly boosts the performance of collaborative retrieval-augmented LLMs, achieving performance gains of 4.23\% to 91.39\% over baseline methods across all tasks.
\end{abstract}


\section{Introduction}
Large Language Models (LLMs)  such as LLaMA and DeepSeek \cite{zhao2023survey,liu2024deepseek,achiam2023gpt,dubey2024llama}, driven by breakthroughs in hardware capabilities and algorithmic innovations, have achieved unprecedented performance in artificial general intelligence. Predominantly deployed on cloud platforms due to their substantial parameter sizes, LLMs necessitate substantial computational resources for inference. However, as LLM applications expand, the increasing demands for real-time responsiveness and strict data privacy requirements have spurred efforts to migrate LLMs from centralized cloud infrastructures to distributed edge environments \cite{lu2024small, zhan2025pice}. 

While cost-optimized edge LLMs with reduced model sizes improve efficiency, they inherently lack contextual awareness and specialized knowledge for domain-specific applications \cite{DBLP:conf/acl/Xiong0LZ24,DBLP:journals/corr/abs-2403-10131,DBLP:journals/corr/abs-2401-08406}. To alleviate these issues, Retrieval-Augmented Generation (RAG) has emerged \cite{lewis2020retrieval,gao2023retrieval,shuster2021retrieval}, which dynamically incorporates external, domain-specific knowledge during inference without modifying model parameters. Specifically, RAG first retrieves the top-$k$ relevant documents from a local vector database and integrates them into the model input, significantly improving contextual understanding and output relevance for subsequent inference.  In contrast to computationally intensive fine-tuning approaches, RAG offers a lightweight, resource-efficient paradigm, which efficiently leverages edge data while preserving its privacy through local processing.

Current research on edge-based RAG systems primarily focuses on single-node optimization, such as on-device inference \cite{jin2024ragcache, ouyang2025adarag} or cloud-assisted frameworks \cite{asai2023self,edge2024local,DBLP:journals/corr/abs-2401-13256}, which assume centralized data storage. However, real-world edge environments are inherently distributed and heterogeneous, rendering single-node RAG systems inadequate due to limited data coverage. As user expectations gradually shift toward real-time, personalized AI interactions, the transition from single-node to multi-node architectures becomes imperative. 
For instance, in smart healthcare \cite{yang2023large,ali2024federated}, multiple hospital nodes deploy specialized LLMs trained on localized private medical knowledge (e.g., for internal medicine or pediatrics) to support AI-assisted diagnostics. This distributed deployment enables dynamic query allocation to the most suitable node based on domain expertise. Furthermore, overlapping domains (e.g., cold symptom diagnosis) can facilitate adaptive workload distribution: during peak demand, such as flu season, queries can be intelligently routed to sub-optimal but capable nodes, alleviating bottlenecks while maintaining load balancing and system responsiveness. Thus, efficient query scheduling is essential, necessitating multidimensional awareness, including domain specialization, resource availability, and current load, to holistically optimize both response quality and latency.

The multi-node RAG paradigm, while promising, introduces key challenges in practical deployment: First, due to privacy concerns, edge nodes typically handle multi-domain data with unknown distributions, posing significant challenges in discerning query intents and dynamically routing requests to domain-specific nodes without prior knowledge of data patterns. Second, query arrivals exhibit spatiotemporal heterogeneity in volume and domain distribution, while edge nodes vary in resource capacities and model capabilities. Static heuristics for query allocation risk node overload, degrading real-time performance. Consequently, balancing load distribution across nodes while preserving LLM output quality remains an open challenge. Third, larger LLMs improve generation quality but increase inference latency. The diversity of model types across resource-constrained nodes creates a combinatorial optimization problem for batch query scheduling, requiring careful trade-offs between quality and latency. 

In this paper, we present \textbf{CoEdge-RAG}, a novel hierarchical scheduling framework designed for multi-node collaboration that jointly optimizes query scheduling, model deployment, and resource allocation to maximize generation quality under strict latency constraints. To enable efficient query-to-corpus matching without prior knowledge of data distributions, we propose an online intent-aware query identification mechanism, which leverages proximal policy optimization (PPO), an advanced deep reinforcement learning technique, to infer query-domain attributions and learn latent matching relationships between queries and edge nodes' domain knowledge. Based on the derived probability distributions, we develop an adaptive inter-node scheduling policy that dynamically routes queries to heterogeneous edge nodes. This allocation adapts to real-time demands by incorporating the estimated node capacity function, ensuring balanced workload distribution and generation quality across the distributed edge network. To further optimize edge node efficiency, we transform the combination optimization of model deployment and resource allocation into an online convex optimization. We then develop an intra-node scheduling strategy to dynamically configure LLM model sizes and memory resources in real-time, complementing our inter-node allocation approach. This fine-grained optimization enables adaptive trade-offs between inference latency and generation quality while preserving computational efficiency. In summary, our hierarchical framework integrates these components to facilitate efficient, data-aware query processing in decentralized edge environments, simultaneously
optimizing both resource utilization and RAG performance. 

The key contributions of this work include:
\begin{itemize}
    \item We focus on the emerging collaborative edge computing paradigm for retrieval-augmented LLMs, boosting LLM deployment towards ubiquitous edge intelligence.
    \item We propose a lightweight online query identification mechanism that establishes latent correlation matching between user queries and local corpora, enhancing retrieval-augmented LLMs at the edge to produce more reliable responses.
    \item Considering heterogeneous edge capacity, we propose a hierarchical scheduling framework to improve response quality within latency limitation. Specifically, centralized inter-node scheduling balances potential generation quality and allocated workload, while distributed intra-node scheduling leverages diverse LLMs to exploit limited edge resources for more efficient LLM serving.
    \item Through a comprehensive evaluation on QA benchmarks with diverse domains,  our system demonstrated high effectiveness and efficiency, achieving 4.23\% to 91.39\% performance improvements over baseline methods.
    
\end{itemize}

\section{Motivation}
\label{motivation}

We conduct several case studies to identify key challenges in developing RAG-enhanced LLMs for collaborative edge computing. Our experimental testbed comprises three edge nodes, each hosting a LLaMA-3B model \cite{dubey2024llama} with a local corpus consisting of 60\% domain-specific data (sports, law, or finance) and 40\% distributed across the other two domains (20\% each). This configuration emulates real-world cross-domain knowledge distribution while maintaining corpus diversity across nodes. Besides, we evaluate output quality using two common metrics: (i) Rouge-L \cite{lin2004rouge} for lexical overlap between generated and reference texts, and (ii) BERTScore \cite{zhang2019bertscore} for semantic similarity using contextual embeddings.

\subsection{Mismatch between Query Allocation and Edge Corpus} 
RAG enhances LLM output quality when edge nodes maintain corpora highly matched to user queries. However, the improvement could diminish significantly under ambiguous corpus distributions across nodes. To verify the correlation between node-query alignment and system efficacy, we simulate 1,500 concurrent queries under three allocation strategies: Random (queries routed without domain awareness), Domain (queries assigned to nodes based on their primary domain focus, e.g., financial queries to the financial node), and Oracle (queries optimally routed to ideal nodes).

\begin{figure}[ht]
  \centering
  \begin{minipage}[t]{0.48\linewidth}
    \centering
    \includegraphics[width=\linewidth, height = 0.65\linewidth]{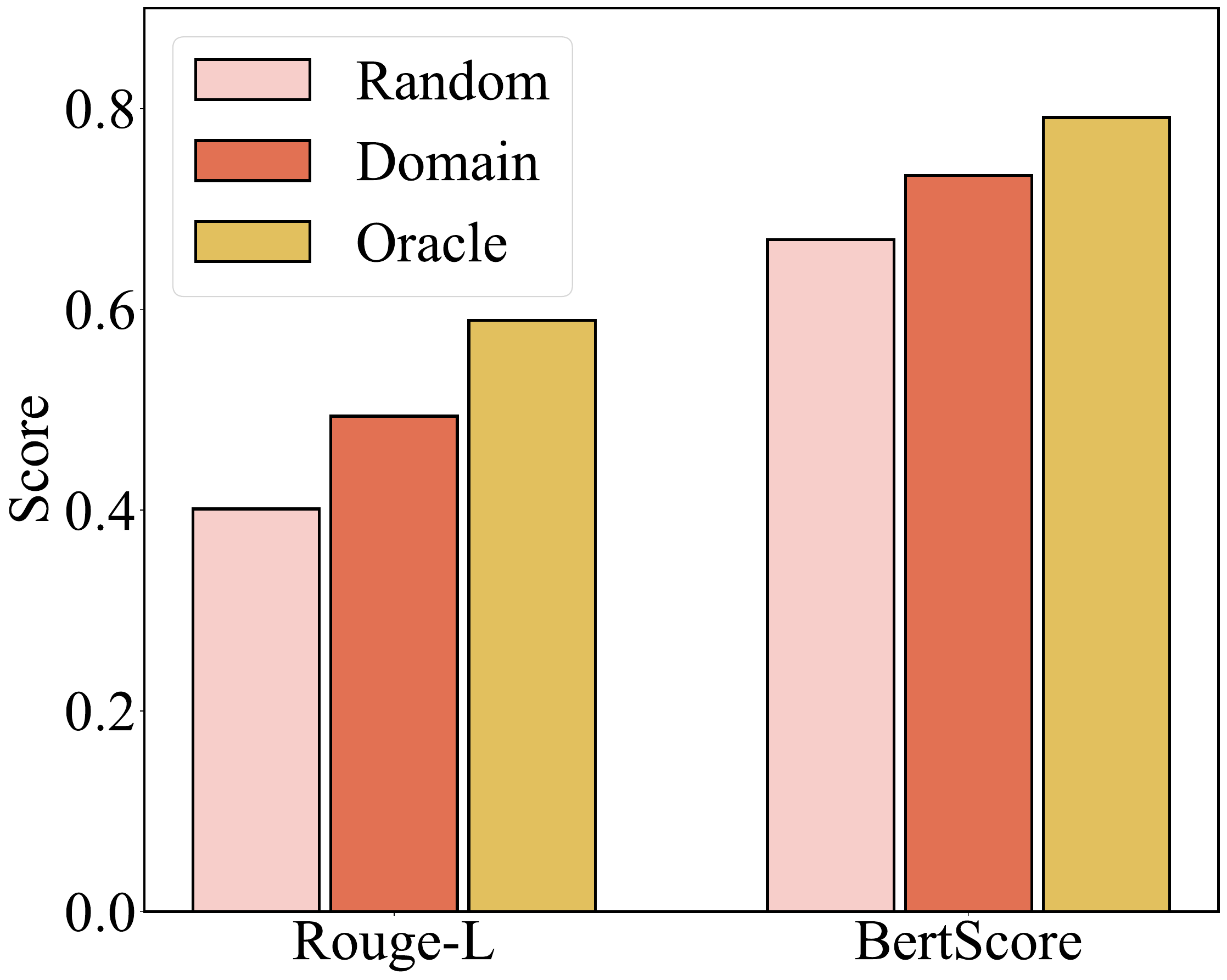}
    \caption{Generation quality comparison.}
    \label{fig:quality_comparison}
  \end{minipage}
  \hfill
  \begin{minipage}[t]{0.48\linewidth}
    \centering
    \includegraphics[width=\linewidth, height = 0.65\linewidth]{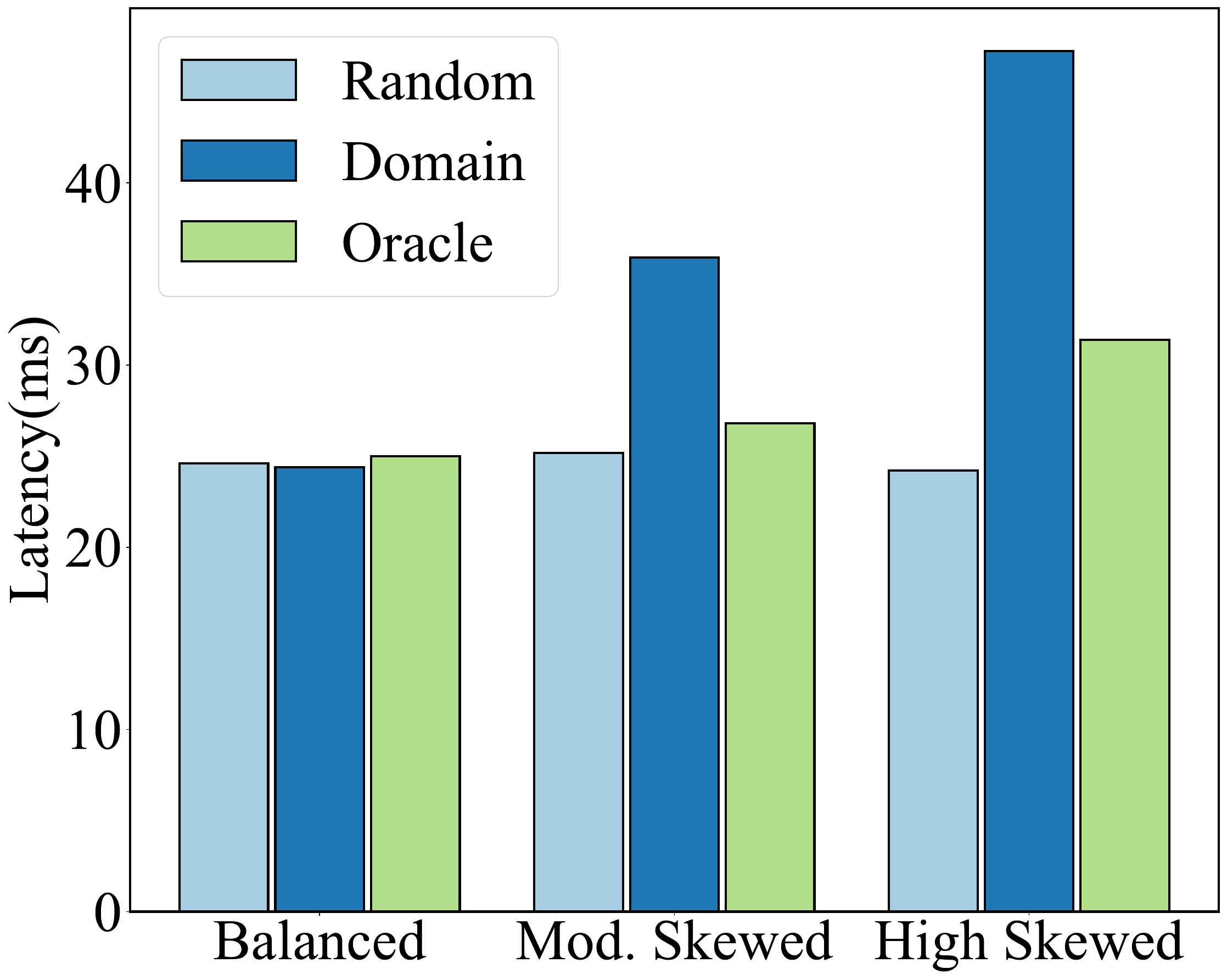}
    \caption{Latency comparison with different skewnesses.}
    \label{fig:latency_comparison}
  \end{minipage}
\end{figure}

As depicted in Fig.~\ref{fig:quality_comparison}, random allocation under opaque corpus conditions degrades performance markedly, exhibiting a 31.9\% reduction in Rouge-L and 15.38\% decline in BERTScore compared to oracle allocation. This is because query patterns fail to match static corpus-node, rendering retrieved documents frequently irrelevant. Domain allocation also exhibits suboptimal performance as it fails to leverage latent cross-domain knowledge from non-primary nodes, thereby constraining the utilization of distributed corpus expertise.

\subsection{Temporal Query Skewness and Latency Degradation} In practice, temporal query skewness --- where user requests concentrate on specific domains during peak periods --- induces resource contention and increases timeout risks, particularly under stringent latency constraints. We measure query latency as end-to-end execution time (from request submission to result delivery). To evaluate the impact, we simulate three skewness patterns across sports/law/finance domains: Balanced (500/500/500 queries), Moderately Skewed (750/375/375, sports-biased), and Highly Skewed (1000/250/250, sports-dominated). As depicted in Fig.~\ref{fig:latency_comparison}, domain-specific allocation suffers severe latency degradation under skewness: moderate skewness and high skewness increase latency by 47.21\% and 93.68\%, respectively, versus balanced scenarios. This stems from fixed routing strategies overloading domain-designated nodes while leaving idle nodes underutilized during peak demand. In contrast, oracle allocation dynamically redistributes queries based on cross-node data relevance, reducing latency by 25.33\%--33.55\% compared to domain allocation. Our findings reveal the necessity of dynamic load-balancing mechanisms that leverage latent inter-domain correlations across nodes, rather than static domain-specific routing, to mitigate timeout risks and optimize resource utilization.

\begin{figure}[t]
  \centering
  \begin{subfigure}{0.48\linewidth}
    {\includegraphics[width=\linewidth, height = 0.65\linewidth]{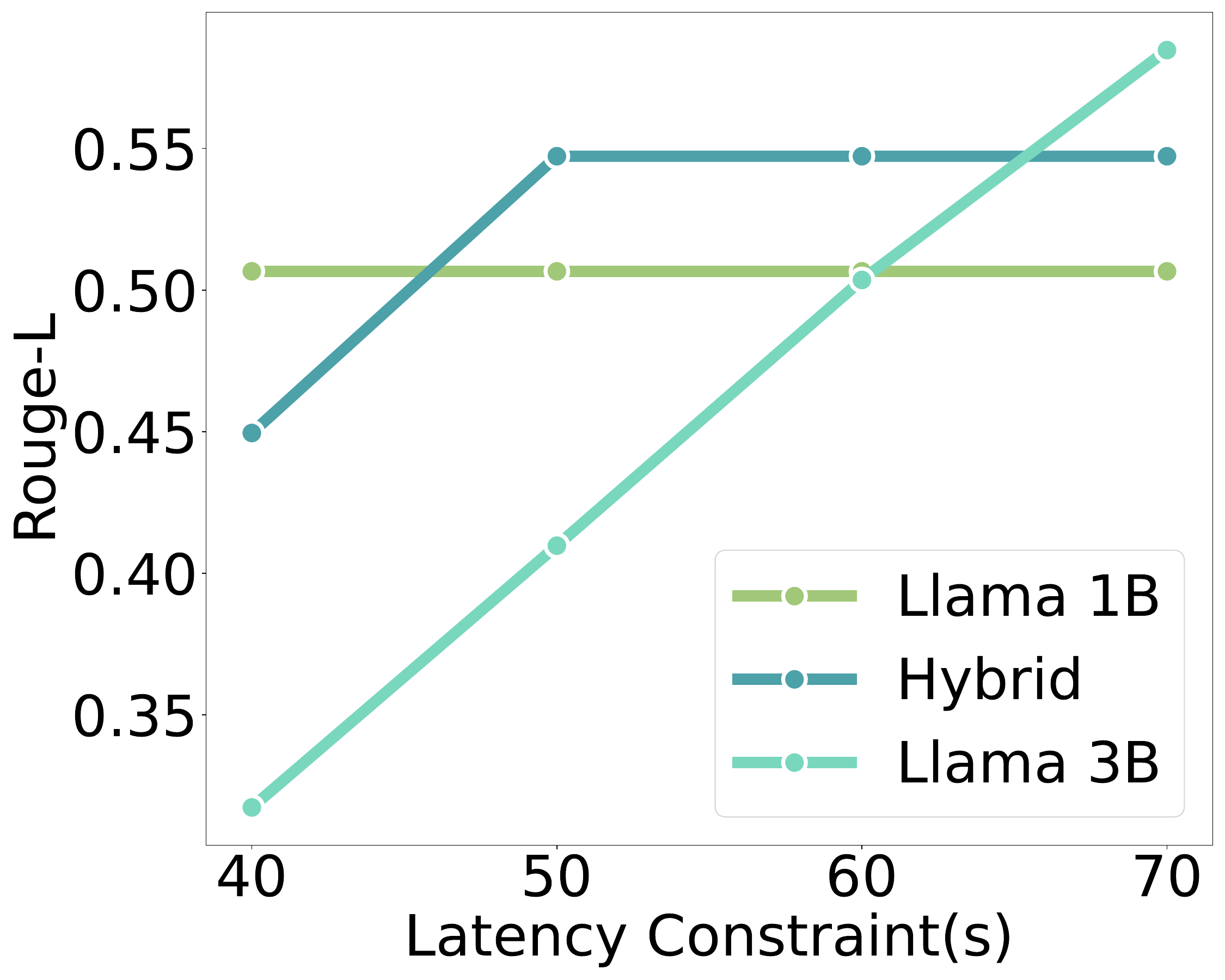}}
    \caption{Generation quality with different model deployments under varying latency constraints.}
    \label{fig:latency_constraint}
  \end{subfigure}
  \hfill
  \begin{subfigure}{0.48\linewidth}
    \centering
    \includegraphics[width=\linewidth, height = 0.65\linewidth]{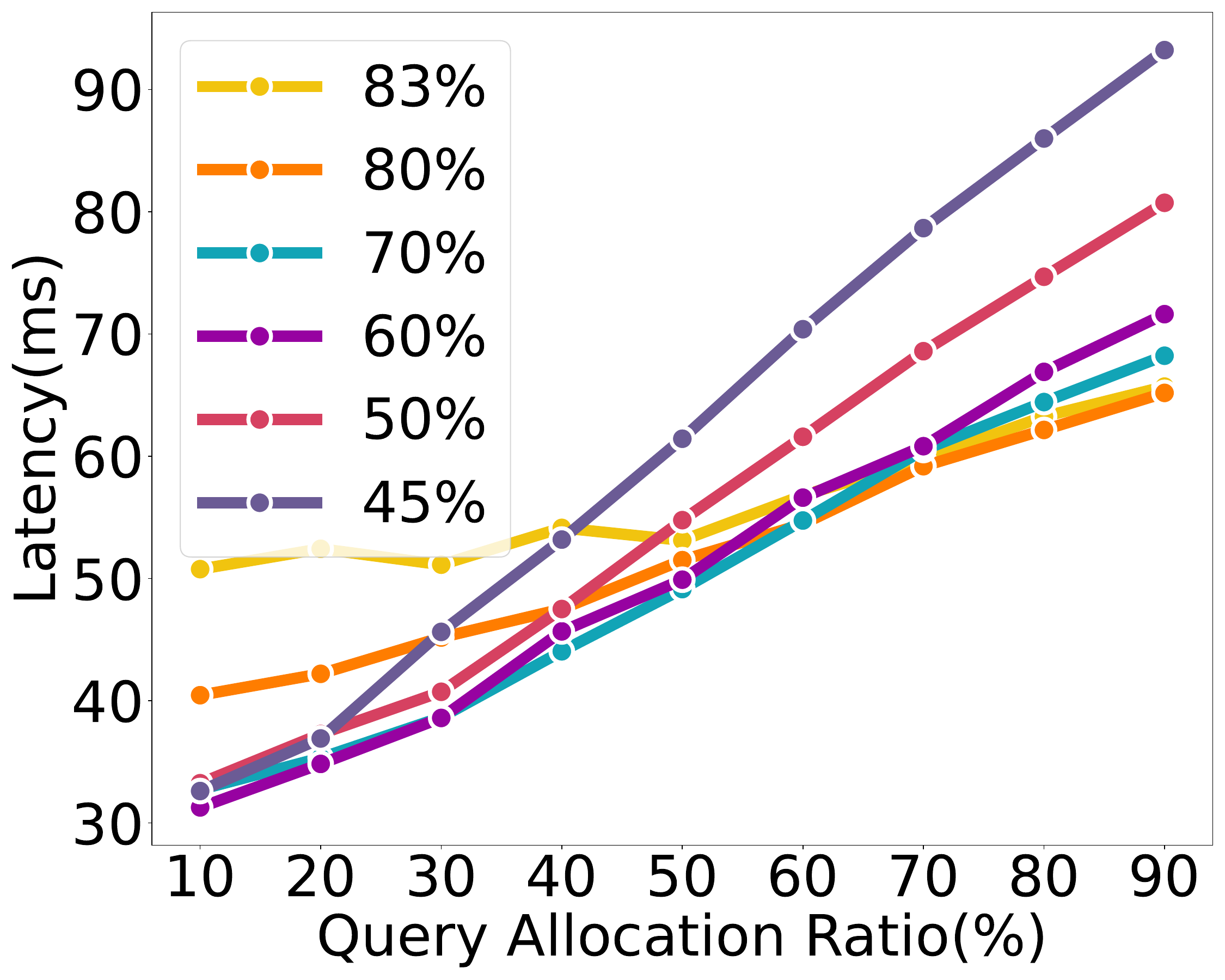}
    \caption{Latency with different resource utilization under varying query ratios.}
    \label{fig:resource_ratio}
  \end{subfigure}
  \caption{Generation quality and latency performances.}
  \label{fig:ratio_comparison}
  \vspace{-15pt}
\end{figure}
\subsection{Quality-Latency Tradeoffs in Model Deployment and Resource Allocation} Resource-constrained edge nodes can strategically deploy multiple LLM variants to balance computational efficiency and output quality. To systematically evaluate this tradeoff, we conducted a comparative analysis of three configurations:
(i) exclusive deployment of LLaMA-1B, (ii) hybrid deployment where LLaMA-1B and LLaMA-3B each processed 50\% of queries, and (iii) exclusive deployment of LLaMA-3B.  
Fig.~\ref{fig:latency_constraint} reveals constraint-driven performance patterns for 1000 requests: \textbf{1) Under strict time constraints ($<$ \SI{50}{s})}: LLaMA-1B achieved a 0.506 Rouge-L score with zero timeout, while hybrid deployment and 3B-exclusive may suffer up to 17.9\% and 45.7\% quality degradation respectively due to timeout-induced interruptions.  \textbf{2) Under relatively relaxed time budget ($>$ \SI{50}{s})}: Rouge-L score of hybrid deployment jumps to 0.547 (8\% improvement over 1B-exclusive), while LLaMA-3B requires more than \SI{70}{s} to fully unleash its potential (0.584 Rouge-L). Thus, a dynamic model deployment is needed based on varying loads and time constraints, rather than rigidly using fixed model deployment, to optimize local service quality.

To demonstrate the necessity of dynamic resource management, we evaluate end-to-end latency under a fixed GPU memory budget while varying resource allocations and query distributions between LLaMA-1B and LLaMA-3B. Fig.~\ref{fig:resource_ratio} illustrates the fraction of GPU memory allocated to LLaMA-3B (45\%--83\%) and the corresponding query allocation ratio (percentage of queries directed to LLaMA-3B). Through adjusting query and resource allocations, we identify two critical scenarios: \textbf{
1) Resource Starvation in Larger Models (45\%--50\% GPU memory for LLaMA-3B):} When LLaMA-3B operates with constrained memory, it becomes a performance bottleneck under increasing query load. Allocating 90\% of queries to LLaMA-3B --- compared to 80\% --- intensifies resource contention, increasing latency by up to 34.1\%. This degradation compromises system responsiveness and risks violating real-time service constraints. \textbf{2) Underutilization of Fast-Response Models (80\%--83\% GPU memory for LLaMA-3B):} Conversely, excessive memory allocation to LLaMA-3B starves LLaMA-1B, even in latency-sensitive workloads where the smaller model should dominate query processing. With insufficient memory, LLaMA-1B suffers high tail latency due to contention, paradoxically increasing system latency by 28.1\%--62.5\% when more queries are routed to it. These results demonstrate a key trade-off: static resource allocation fails to accommodate dynamic query routing patterns, whereas efficient edge inference requires the co-design of query scheduling and memory allocation with runtime rebalancing to satisfy real-time service requirements.

\section{Design of Collaborative LLMs Serving}
CoEdge-RAG aims at maximizing generation quality while maintaining satisfactory latency requirements. Key insights are as follows: (i) in terms of generation quality, queries associated with specific attributes (e.g., subject) are more likely to match local corpora data at specific edge nodes, thus enhancing the RAG performance of the edge-deployed LLMs; (ii) in terms of response latency, efficient workload balancing across multiple edge nodes mitigates straggler effects, where a single node is overwhelmed by excessive queries, thus reducing overall serving time. 

\begin{figure}[t]
    \centering
    \includegraphics[width=\linewidth]{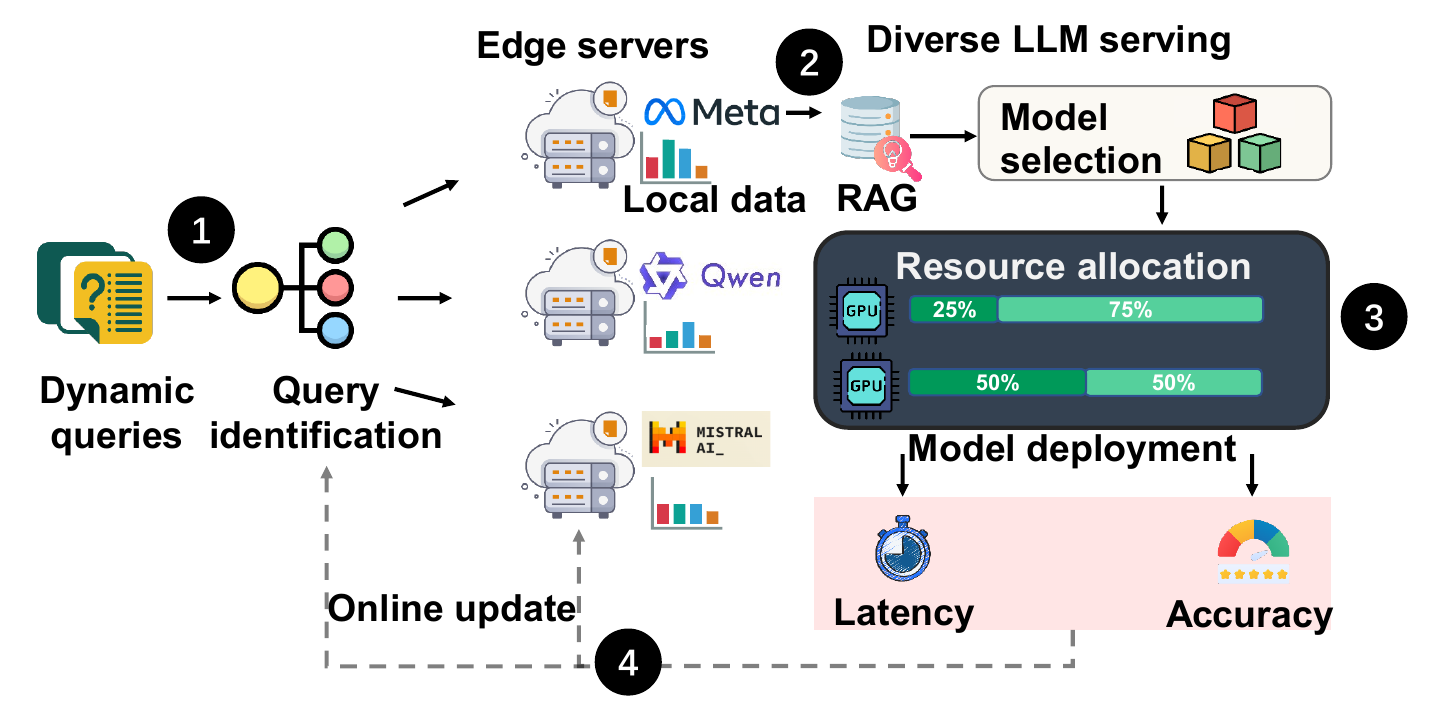}
    \caption{An illustration of CoEdge-RAG.}
    \label{fig:illustration}
    \vspace{-20pt}
\end{figure}

The CoEdge-RAG workflow, illustrated in Fig.~\ref{fig:illustration}, operates as follows: A global coordinator \scalebox{0.9}{$\blackcircle{1}$} first tokenizes and encodes incoming queries into embedding vectors, and computes corresponding probability distributions via an online query identifier. Such probability vectors indicate matching degrees between queries and edge nodes, assisting the coordinator to route queries to designated nodes. Upon receipt, nodes \scalebox{0.9}{$\blackcircle{2}$} retrieve the top-$k$ semantically relevant local documents via vector similarity search, augmenting queries with retrieved contexts to form enriched prompts for subsequent LLM processing. Based on query volumes and latency requirements, each node \scalebox{0.9}{$\blackcircle{3}$} dynamically selects optimal LLMs from its model pool for per-GPU deployment and allocates suitable resource and query proportions to each model. LLM outputs are assessed via human or automated evaluators, generating quality metrics (e.g., Rouge, BertScore). These metrics \scalebox{0.9}{$\blackcircle{4}$} are used to update the online query identifier mechanism, establishing an iterative optimization loop that strengthens data-characteristic alignment between queries and edge nodes.

Note that our work focuses on single-document queries, wherein the required information can be derived from a single source document. It aligns with standard RAG baselines \cite{asai2023self,seemakhupt2024edgerag}, enabling a clear and focused evaluation of the performance gains directly attributable to our novel hierarchical scheduling framework. To address more complex queries, our system can be easily extended with a preprocessing module that decomposes them into sub-queries \cite{sawarkar2024blended}. The scheduler would then allocate these sub-queries across the diverse edge nodes, thereby preserving its core optimization benefits while broadening applicability.

\subsection{Query Identification and Scheduling Optimization}
\label{Query Identification and Scheduling Optimization}
\textbf{Query Identification:} To efficiently characterize dynamic edge environments, a slot structure is adopted to describe the long-term CoEdge-RAG procedure. At each time slot $t\in \mathcal{T}$, a bunch of queries with numbers $B^t$ must be served by a collection of edge nodes $\mathcal{N}$. To leverage the characteristics of each query, an online identifier is adopted to establish initial latent correlation matching between user queries and each node corpus. Specifically, each query $i$ is first processed by the carefully designed encoder to a semantically rich text embedding. Then the online query identifier will take it as input, and generate a probability vector $\mathbf{s}^t_{i} = (s^t_{i1},s^t_{i2},..., s^t_{iN})$ to quantify the matching degree between query $i$ and each edge node, where entries in the vector hold $\underset{n\in\mathcal{N}}\sum \mathbf{s}^t_{in}=1$. Naturally, the set of all probability vectors for queries arriving at slot $t$ is represented by $\mathcal{S}^t$ = \{$\mathbf{s}_{1}^t$, $\mathbf{s}_{2}^t, \dots,\mathbf{s}_{B^t}^t$\}. 
These probability vectors facilitate informed decision optimization in the following allocation scheduling process and determine the final allocation node for each query.

\textbf{Scheduling Optimization:} Subsequently, each edge node can utilize its constrained resources by deploying suitable LLMs to handle these queries. As outlined in Section \ref{motivation}, edge nodes typically host diverse model series (e.g., LLaMA and Qwen), each maintaining a dedicated model pool $\mathcal{M}_n$ containing parameter-scaled variants for GPU deployment. Even though an edge node is equipped with multiple GPUs, to maximize LLM efficiency, we implement a single-GPU allocation strategy guided by two principles: (i) aligning with edge resource limitation, we prioritize smaller-scale LLMs (e.g., 1B--8B parameters), which can be efficiently supported at a single GPU. (ii) Deploying an LLM across multiple GPUs incurs inter-GPU communication and memory synchronization costs, increasing system complexity and degrading inference latency compared to single-GPU execution. In this way, for model $m\in\mathcal{M}_n$ on GPU $k$ of node $n$, inference latency is governed by the query load $p_{mnk}^t$ (proportion of allocated queries) and resource allocation $R_{mnk}^t$ (proportion of GPU resources), subject to the constraint $ \sum\limits_{n}\sum\limits_{k} \sum\limits_{m} p_{mnk}^t = 1$. These scheduling decisions are dynamically adjusted in real-time to system dynamics, enabling adaptive optimization of model placements and resource partitioning across the edge hierarchy.

\subsection{Performance Metrics}
\label{3.2}
\textbf{Latency:} In CoEdge-RAG, the time of the RAG inference service mainly includes the vector search time, the model processing time, and the time overhead induced by GPU resource reallocation. The vector search time of retrieving documents, denoted as $TS_n^t$, can be measured in real-time before performing LLM inference at each node $n$. The processing latency is determined by the query proportion $p^t_{mnk}$ and allocated resources $R^t_{mnk}$, expressed as  $L^t_{mnk}(p^t_{mnk}B^t, R_{mnk}^t)$. Regarding GPU resource reallocation, time overhead may arise due to dynamic adjustments in model deployment strategies under varying query loads. Specifically, adapting to workload fluctuations may involve (i) loading previously undeployed models, (ii) unloading existing models, or (iii) adjusting resource allocations for currently deployed models.

Based on empirical measurements, the unloading operation typically incurs negligible overhead (a few hundred milliseconds) compared to model loading or resource reallocation. To formally characterize these behaviors, we first define the deployment status of model $m$ in the model pool $\mathcal{M}_n$ on GPU $k$ at time slot $t$ as $d^t_{mnk} \in \{0,1\}$, where $d^t_{mnk}=1$ indicates active deployment and $d^t_{mnk}=0$ otherwise. The unloading state of model $m$ is then derived as:
\begin{equation}
ULD_{mnk}^t = (1 - d_{mnk}^t) \cdot d_{mnk}^{t-1},
\end{equation}
where $ULD_{mnk}^t = 1$  signifies an ongoing unloading operation at time slot $t$, while $ULD_{mnk}^t = 0$ indicates no unloading occurs. This formulation explicitly isolates unloading events, which is critical because although unloading induces changes in resource allocation($R^t_{mnk}$), its time overhead is nearly negligible. Thus, the primary sources of reconfiguration latency are: (i) loading previously undeployed models $(d_{mnk}^{t} = 1 \wedge d_{mnk}^{t-1} = 0)$, and (ii) adjusting resources for persistent models $(R^t_{mnk} \neq R^{t-1}_{mnk} \wedge ULD_{mnk}^t=0)$.



Moreover, to prevent resource contention during model initialization, we enforce serialized loading operations on each GPU. Consequently, the total loading time for GPU $k$ during time slot $t$ becomes the sum of individual loading times for all models requiring deployment. Let $l_m$ represent the loading time of the model $m$. The aggregate loading time on GPU $k$ at time slot $t$ is then given by:
\begin{equation}
    \label{eq:loadtime}
   TL_{k}^{t} = \sum\limits_{m} \mathbbm{1}_{R_{mnk}^t\neq R_{mnk}^{t-1}} \cdot (1-ULD_{mnk}^t) \cdot l_m,
\end{equation}
where the indicator function  $\mathbbm{1}_{R_{mnk}^t\neq R_{mnk}^{t-1}}$
evaluates to 1 when the memory allocation for model $m$ has changed between time slots, and 0 otherwise. 

\textbf{Generation Quality:} Let $Q_{mnk}^t(p_{mnk}^tB^t, R_{mnk}^t,\mathcal{S}^t)$ represent the average generation quality of model $m$ in node $n$ within time slot $t$, whose value depends primarily on the number of queries $p_{mnk}^tB^t$ it receives, resource proportion $R_{mnk}^t$ and query identification vectors $\mathcal{S}^t$. Additionally, if model $m$ exceeds its relevant latency requirements while processing $p_{mnk}^tB^t$ queries, the portion of queries that exceed the requirements will be considered invalid.

\subsection{Problem Formulation}
We formulate the \textbf{CoEdge-RAG} optimization problem as maximizing overall generation quality under strict latency constraints across resource-constrained edge nodes:
\begin{gather}
\max \sum\limits_{t}\sum\limits_{m,n,k} p^t_{mnk} Q_{mnk}^t(p_{mnk}^tB^t, R_{mnk}^t, \mathcal{S}^t)  \\
\text{s.t. } \max\limits_{m,n,k} \left[L^t_{mnk}(p^t_{mnk}B^t,R_{mnk}^t) + TL_{k}^t\right] \leq L^t-TS_n^t,\label{eq:intra2}\forall t, \\
\sum\limits_{m} R_{mnk}^t \leq R_{k}, \forall n,k,t, \label{eq:intra3} \\
R_{mnk}^{t} \geq {d_{mnk}^t r_{m}}, \forall m,n,k,t, \label{eq:intra4}\\
R_{mnk}^{t} \leq d_{mnk}^t, \forall m,n,k,t, \label{eq:intra5}\\
\sum\limits_{n}\sum\limits_{k}\sum\limits_{m} p^t_{mnk} = 1, \forall t,\label{eq:intra6}
\end{gather}
where $L^t$ represents the latency requirement specified in the service-level objective (SLO), which may vary temporally due to dynamic factors such as fluctuating user demands, variable network conditions, or shifts in service priorities (e.g., stricter latency during peak hours).
$TS_{n}^t$ (at time slot $t$) denotes the vector search time for queries, 
$R_{k}$ denotes the GPU constraints of GPU $k$ in node $n$ and $r_{m}$ denotes minimum model-startup memory utilization of model $m$. Eq.~(\ref{eq:intra2}) ensures that every model allocated queries and resources meets the latency requirement $L^t$.
Eq.~(\ref{eq:intra3}) enforces GPU memory constraints in each card.  Eq.~(\ref{eq:intra4}) ensures the minimum memory is allocated to each deployed model. Eq.~(\ref{eq:intra5}) forces resources to zero for undeployed models. Eq.~(\ref{eq:intra6}) ensures that queries can be dispatched to models for processing. Unfortunately, solving the above problem is difficult due to the following challenges: ($\mathcal{C}_1$) Accurate estimation of probability vectors $\mathcal{S}^t$ is hindered by the heterogeneity of edge corpora, unknown data distributions, and the high dimensionality of query features; ($\mathcal{C}_2$) The generation quality $Q_{mnk}^t(\cdot)$ and processing latency $L^t_{mnk}(\cdot)$ lack closed-form expressions, complicating the optimization of scheduling policies; ($\mathcal{C}_3$) CoEdge-RAG must dynamically adjust sequential scheduling decisions to accommodate evolving user queries and heterogeneous edge environments without prior knowledge of system dynamics.

\section{Hierarchical Scheduling Optimization}
In this section, we propose a hierarchical scheduling framework, \textbf{CoEdge-RAG}, explicitly designed to address the scalability and heterogeneity challenges inherent to collaborative edge systems. 

\subsection{Data-aware Online Query Identification}
High-quality retrieval data is crucial for enhancing the effectiveness of RAG-based LLMs, particularly for domain-specific user queries. However, privacy-preserving issues typically obscure the detailed metadata about corpus composition and distribution across edge nodes. To address challenge $\mathcal{C}_1$, we develop an online query identification mechanism capable of inferring corpus relevance across nodes without exposing sensitive distribution metadata.

To model the latent correlation between queries and edge nodes, each query $i$ is first encoded into a semantically rich embedding vector $e_i^t$ using a pre-trained text encoder\footnote{https://huggingface.co/BAAI/bge-base-en-v1.5}. Such an embedding well captures domain-specific attributes and informs a routing policy network to generate probability distribution $s_i^t$ (defined in Section \ref{Query Identification and Scheduling Optimization}), representing node-specific relevance scores. Since response quality implicitly reflects query-document alignment, we implement a feedback-driven mechanism that optimizes the policy network through iterative refinement. For each query $i$, the edge-generated response $GEN_i$ is evaluated against a powerful cloud-LLM (e.g., DeepSeek-V3) reference $REF_i$, which serves as a quality benchmark for contextually aligned outputs~\cite{ouyang2025adarag}. Response quality is assessed using two complementary metrics: \textbf{Lexical Alignment}: ROUGE-L~\cite{lin2004rouge}  computes the normalized longest common subsequence (LCS) between $REF_i$ and $GEN_i$:
\begin{equation*}
   f_{i,R}^t = 
   \frac{LCS(REF_i,GEN_i)}{max(len(REF_i),len(GEN_i))}.
\end{equation*}
\textbf{Semantic Consistency}: BERTScore~\cite{zhang2019bertscore} measures the harmonic mean of the maximum cosine similarities in both directions (i.e., from generated to reference tokens and vice versa). The precision and recall are defined as follows:
\begin{align*}
    \mathrm{Prec}(\cdot) &= \frac{1}{|GEN_i|} \sum_{k=1}^{|GEN_i|} \max_j \mathrm{sim}(E(GEN_i)_k, E(REF_i)_j), \\
    \mathrm{Rec}(\cdot) &= \frac{1}{|REF_i|} \sum_{j=1}^{|REF_i|} \max_k \mathrm{sim}(E(REF_i)_j, E(GEN_i)_k),
\end{align*}
where $E(REF_i)$ and $E(GEN_i)$ represent the token-level embeddings of the reference and generated responses, respectively. The final BERTScore is then computed as:
\begin{equation*}
f_{i,B}^t = \frac{2 \cdot \mathrm{Prec}(GEN_i, REF_i) \cdot \mathrm{Rec}(REF_i, GEN_i)}{\mathrm{Prec}(GEN_i, REF_i) + \mathrm{Rec}(REF_i, GEN_i)}.
\end{equation*}

These complementary metrics (i.e., $f_{i,R}^t$ for lexical alignment and $f_{i,B}^t$ for semantic fidelity) are combined into a composite quality score:
\begin{equation}
    f_i^t = \alpha_1 \cdot f_{i,R}^t  + \alpha_2 \cdot f_{i,B}^t,
\end{equation}
where $\alpha_1$ and $\alpha_2$ are their weight factors. The rationale for using $f_i^t$ as a proxy for node relevance is grounded in the RAG paradigm's fundamental principle: higher response quality indicates stronger alignment between the query and the retrieved corpus. Based on the definition, $f_i^t$ quantifies output performance on the assigned node, reflecting its latent relevance. This feedback implicitly estimates query-corpus alignment while evaluating the efficacy of edge-deployed models, thereby guiding subsequent policy network updates.

For policy network optimization, we employ Proximal Policy Optimization (PPO) \cite{schulman2017proximal}, an advanced reinforcement learning algorithm, to enhance our online query identification framework.  In order to accommodate stochastic and non-interfering query streams with high-dimensional semantic features, we devise a streamlined architecture that eliminates the value function network while maintaining a policy-only structure. This design choice yields dual benefits: (i) substantial reduction in computational complexity, and (ii) improved processing efficiency. Empirical evaluations demonstrate an average inference latency of \SI{0.02}{ms} per query, satisfying stringent real-time system constraints.

In the absence of a critic network, we formulate a standardized feedback metric as the reward signal within each batch:

\begin{equation}
    \Bar{f}_i^t = \frac{f_i^t - \mu}{\sigma + c},
    \label{std}
\end{equation}
where $\mu$ and $\sigma$ denote the batch-wise mean and standard deviation of feedback values, respectively, and $c = 10^{-8}$ is a small constant to avoid division by zero. This operation ensures stable gradient magnitudes across varying query workloads. In this way, the policy parameters $\theta$ are then optimized through a modified PPO objective:
\begin{equation}
    L_f =\mathbb{E}_t \left[ \min (\rho_i^t \Bar{f}_i^t,\operatorname{clip}(\rho_i^t, 1-\epsilon, 1+\epsilon) \Bar{f}_i^t) \right] + \beta H(\pi_\theta),
\end{equation}
where $\pi_\theta$ and $\pi_{\theta_{old}}$ denote the updated policy network and the fixed old policy network, respectively. The importance sampling ratio is defined as
$\rho_i^t = \pi_\theta(e_i^t) /\pi_{\theta_{old}}(e_i^t)$.
Using this ratio and the CLIP mechanism, PPO maintains training stability by preventing drastic policy deviations.
This approach also facilitates efficient batch reuse for iterative policy refinement, mitigating performance instability from abrupt updates. These properties align with edge computing requirements for sample efficiency, low computational overhead, and fast adaptation to dynamic data, ensuring robust performance in resource-constrained settings. Additionally, the entropy term $\beta H(\pi_\theta)$ encourages exploration, preventing premature convergence. In summary, the modified objective $L_f$ achieves an efficient trade-off between two critical objectives: (i) feedback-guided refinement through intermediate node LLM outputs to establish query-document associations, and (ii) stochastic exploration via policy entropy to ensure diverse node selection.

To minimize training frequency while preserving system stability, we employ a memory buffer that accumulates historical feedback and triggers batched policy updates only when the accumulated queries exceed a predetermined threshold. This threshold is statically configured based on the average query load observed over an extended time horizon, ensuring updates are neither too frequent to incur unnecessary computational overhead nor too sparse to compromise adaptability. By decoupling updates from transient query fluctuations and instead aligning them with statistically representative workloads, this approach effectively achieves robust training stability while reducing the number of network updates. The update process is designed for efficiency, typically completing within \SI{30}{ms} per 1000 queries, thereby achieving a favorable balance between computational resource conservation and responsiveness in dynamic edge environments.

\subsection{Loading-balancing Inter-node Scheduling}
\label{4.2}

While online query identification establishes initial query-node mappings, it neglects computational heterogeneity and inherent node capacity constraints. Consequently, it may introduce critical bottlenecks when bunches of query workloads are concentrated on a specific domain, inducing huge processing delays and service degradation on an overwhelmed edge node. Prior to addressing $\mathcal{C}_2$ and $\mathcal{C}_3$, we propose an inter-node scheduling mechanism that dynamically optimizes the query allocation proportion $p_n^t$ for each node $n$ (where $p_n^t:=\sum\limits_{k} \sum\limits_{m} p_{mnk}^t$), thus balancing the trade-off between generation quality and workload distribution via two objectives: (i) prioritizing allocation to nodes with probability scores $\mathcal{S}^t$ and (ii) balancing allocation based on nodes' capacity. Once the query proportion $p_n^t$ of each node is determined, the joint scheduling problem in collaborative edge computing decomposes into independent intra-scheduling subproblems (Section \ref{intra-paragraph}), thereby largely reducing computational complexity.

The core principle of inter-node scheduling is to determine the maximum query capacity of each node while adhering to predefined latency constraints. This ensures efficient query distribution, preventing node overload. Additionally, it promotes balanced data exploration across nodes, enhancing performance under high-load scenarios.

During the initialization phase, the system autonomously profiles each node's processing capacity through controlled query bursts across progressively relaxed latency requirements $L$. The latency parameter $L$ is systematically varied from \SI{5}{s} to \SI{60}{s} in \SI{5}{s} increments, revealing the node-specific sensitivity to time constraints. Beginning with $L =$ \SI{5}{s}, the system progressively increases the query load on the target node until the query drop rate surpasses a predefined threshold (e.g., 1\%). This establishes the maximum sustainable throughput $E_{n,5}$  for node $n$ under the \SI{5}{s} latency constraint. For subsequent latency levels, the system initializes the query distribution volume to node $n$ as ($L/5$) $\cdot$ $E_{n,5}$.  If the drop rate remains below the threshold, the load is incrementally increased by $E_{n,5}$ until the threshold is exceeded, at which point the current throughput is recorded as $E_{n,L}$. Building upon this empirical characterization, the system applies linear regression for simplicity to derive a node-specific capacity function $C_n(L^t)$,  which models the maximum sustainable query throughput for node $n$ while satisfying an arbitrary latency requirement $L^t$:
\begin{equation}
    C_n(L^t) = k_n L^t + b_n,
\label{Cn}
\end{equation}
where $k_n$ and $b_n$ are node-specific coefficients obtained through the offline fitting process. This dynamic capacity estimation guides runtime-adaptive query allocation.


Upon completing initialization, the system will proceed to the subsequent runtime execution phase. As shown in Algorithm  \ref{alg:query_allocation}, the inter-node scheduling algorithm first checks whether the total query volume exceeds the cumulative capacity of all nodes. If so, it proportionally scales up each node's capacity limit temporarily.
The algorithm then proceeds by initializing $q_n$(the number of queries) assigned to each node to zero, and the allocated node index $a_i^t$ of each query $i$ to -1 (indicating unassigned status). Each query $i$ is subsequently assigned to a node $a_i^t$ sampled from the probability vector $s_i^t$. Following this, the system performs a capacity-aware validation: if the node's current load $q_{a_i^t} > C_{a_i^t}(L^t)$, the inter-node scheduling algorithm would reselect an alternative node from the subset $A_i$ of nodes with residual capacity using the renormalized probabilities $\tilde{\mathbf{s}}_i^t$. 
This design ensures each node processes a proper number of queries while maintaining probability-driven load balancing, with the capacity adjustment mechanism providing elasticity to handle instantaneous query surges. 
Finally, the actual allocation proportion $p_j^t = \frac{q_j}{B^t}$ is calculated for node $j$, which effectively decomposes the global query scheduling and processing problem into a set of distributed local subproblems at individual nodes.
\begin{algorithm}[!t]
\caption{Inter-Node Scheduling Algorithm}
\label{alg:query_allocation}

\begin{algorithmic}[1]
\State \textbf{Input:} Total number of queries $B^t$, probability vectors $\mathcal{S}^t$ = \{$\mathbf{s}_{1}^t$, $\mathbf{s}_{2}^t$,..., $\mathbf{s}_{B^t}^{t}$\}, maximum number of queries each node can process $C_n$.
\State \textbf{Output:} Allocation node $a_i^t$ for each query $i$ and query proportion $p_j^t$ for each node $j$
\State Initialize $a_i^t=-1$ for each query $i$ 
\State Initialize $q_j=0$ for each node $j$     
\If{$B^t > \sum\limits_{n} C_n$}
\State $\Bar{C} \gets \sum\limits_{n} C_n$
\State $ C_n \gets C_n + \frac{C_n}{\Bar{C}}(B^t - \Bar{C}) $
\EndIf

\For {each query $i$}
\State $a_i^t \gets \Call{RandomSample}{\mathbf{s}_i^t}$

\If{$q_{a_i^t} \geq C_{a_i^t}$} \Comment{Check if node exceeds capacity}
    \State $A_i \gets \{j \mid q_j < C_j, \forall j \in \mathcal{N}\}$ \Comment{Available nodes with residual capacity}
    \State $\tilde{\mathbf{s}}_i^t \gets \text{Normalize}(\mathbf{s}_i^t[j] \text{ for } j \in A_i)$ 
    \State $a_i^t \gets \Call{RandomSample}{\tilde{\mathbf{s}}_i^t}$ \Comment{Reassign with normalized probabilities}
\EndIf
\State $q_{a_i^t} \gets q_{a_i^t} + 1$ \Comment{Update node's query count}
\EndFor
\State $p_j^t = \frac{q_j}{B^t}$ for each node $j$
\end{algorithmic}
\end{algorithm}
\subsection{Adaptive Intra-node Scheduling}
\label{intra-paragraph}
To achieve fine-grained RAG performance optimization, we refine intra-node scheduling policies, focusing on two key aspects: \textbf{dynamic model resource allocation} and \textbf{intra-node query allocation}. The resource allocation mechanism critically balances generation quality and operational efficiency, as larger models improve quality but increase computational latency. Under strict SLO latency constraints, dynamically optimizing $p_{mnk}^t$ and $R_{mnk}^t$ is essential for efficient LLM deployment. To address challenges $\mathcal{C}_2$ and $\mathcal{C}_3$, the core idea is to approximate generation quality and latency using convex functions, then applying online convex optimization techniques \cite{YuNW17} to derive efficient scheduling decisions without future system information.

As outlined in Section~\ref{3.2}, we employ a latency prediction function $L^t_{mnk}$ for each model $m$,
which estimates processing time given query load and resource allocation. However, as noted in $\mathcal{C}_2$, the absence of closed-form expressions for $L^t_{mnk}$ precludes direct convex optimization. To bridge this gap, we empirically evaluate four candidate convex functions (linear, quadratic, exponential, cubic) by measuring latency across varying resource allocations and query loads. Using rigorous curve-fitting, we derive latency prediction models and evaluate their accuracy via Root Mean Square Error (RMSE).

As demonstrated in Table~\ref{tab:rmse_comparison_en} for the LLaMA model series (DomainQA dataset; see Section 5.1), the quadratic function achieves superior empirical fidelity, with normalized RMSE (NRMSE) values of 2.58\% (LLaMA-1B), 6\% (LLaMA-3B), and 1.87\% (LLaMA-8B) --- striking an optimal balance between computational tractability and prediction error. Consequently, we adopt the quadratic approximation for real-time latency prediction:

\begin{equation}
    \widetilde{L}_{mnk}^t ={(ap^t_{mnk} B^t - b R_{mnk}^t)}^2 + c p^t_{mnk}B^t + d R_{mnk}^t + e + \Delta T,
\end{equation}
where $a, b, c, d, e$ are fitted parameters derived from empirical data, and $\Delta T$ represents a systematic latency offset incorporated to enhance robustness against unmodeled real-world perturbations, such as network delays. Particularly, we adopt fixed-length document chunks and a fixed number of retrieved documents, which significantly simplifies the modeling process. Since user queries are generally short, the input length is dominated by the number of retrieved documents. This allows inference latency to scale approximately linearly with the retrieval count, enabling the latency predictor to generalize across different retrieval quantities via simple proportional scaling, eliminating the need for repeated modeling \cite{ouyang2025adarag, jin2024ragcache}.

\begin{table}[htbp]
    \centering
    \caption{Comparison of RMSE across different functions and models.}
    {\renewcommand{\arraystretch}{1.2}
    \begin{tabular}{|l|c|c|c|c|}
        \hline
        Model & Linear  & Quadratic  & Exponential  & Cubic  \\
        \hline
        LLaMA-1B & 1.449 & 1.141 & 1.130 & 1.118 \\
        \hline
        LLaMA-3B & 1.183 & 0.674 & 0.839 & 0.936 \\
        \hline
        LLaMA-8B & 2.289 & 1.033 & 2.136 & 2.402 \\
        \hline
    \end{tabular}}
    \label{tab:rmse_comparison_en}
\end{table}

Likewise, the quality of model outputs intrinsically depends on the relevance of retrieved contexts, making an accurate a priori assessment particularly challenging. Transient, request-level quality estimations are not only computationally costly but also prone to misrepresenting a model’s true capabilities. This distortion arises because poor retrieval --- an upstream failure (e.g., query identification) --- artificially suppresses output quality, causing even highly capable models to appear deficient.
To isolate effective generative performance from retrieval noise, we employ an offline evaluation framework. Specifically, each node $n$ conducts batch testing using locally deployed models $m$ and node-specific data. Crucially, to control for retrieval variability, models are provided with queries paired with ground-truth context documents. This setup functions as a controlled “open-book” examination, ensuring performance differences reflect only the model's reasoning ability under ideal context conditions rather than stochastic retrieval outcomes. From this evaluation, we derive a static quality score $Q_{mn}$, representing the average intrinsic performance of model $m$ on the data distribution of node $n$ under optimal context availability. This score inherently captures both the model's generative capability and node-specific data characteristics. For scheduling purposes, $Q_{mn}$ serves as a robust heuristic: in the absence of certainty about retrieval success for an incoming query, the node assigns it to the model most likely to utilize relevant context effectively. Thus, we reduce the complex, dynamic quality function $Q_{mnk}^t(p_{mnk}^tB^t,R_{mnk}^t, \mathcal{S}^t)$ to a constant approximation $Q_{mn}$. Additionally, empirical results (Table~\ref{tab:intra_performance} and Fig.~\ref{fig:proportion}) confirm the efficacy and robustness of this approach, showing that $Q_{mn}$ provides a reliable basis for model selection and scheduling.

Meanwhile, the indicator function in Eq.~(\ref{eq:loadtime}) is also non-solvable in the convex optimization problem. Therefore, we introduce a binary variable $RC_{mnk}^t\in \{0,1\}$ to track resource change in time slot $t$. $RC_{mnk}^t$ satisfies the following constraint:
\begin{gather}
R_{mnk}^{t-1} - R_{mnk}^{t} \leq \varepsilon_{1} + M \cdot RC_{mnk}^t, \\
R_{mnk}^{t} - R_{mnk}^{t-1} \leq \varepsilon_{1} + M \cdot RC_{mnk}^t, \\
R_{mnk}^{t-1} - R_{mnk}^{t} \geq \varepsilon_{1} - M \cdot (1 - RC_{mnk}^t), \\
R_{mnk}^{t} - R_{mnk}^{t-1} \geq \varepsilon_{1} - M \cdot (1 - RC_{mnk}^t),
\end{gather}
where $\varepsilon_{1}$ is a small threshold ensuring only significant resource changes are considered, and $M$ is a sufficiently large constant. 
These constraints ensure that the binary tracking variable activates only when the resource change between consecutive time slots exceeds a small predefined threshold. 
After this transformation, $TL_{k}^{t}$ can be approximated as:
\begin{equation}
    \label{loadtime_tran1}
   \overline{TL}_{k}^{t} = \sum\limits_{m} RC_{mnk}^t \cdot (1-ULD_{mnk}^t) \cdot l_m.
\end{equation}


However, in Eq.~(\ref{loadtime_tran1}), variable multiplication remains impermissible in convex optimization. Therefore, we further simplify the expression. As established in Section \ref{3.2}, there exist three distinct resource adjustment strategies, each resulting in $RC_{mnk}^t = 1$. Crucially, whenever $RC_{mnk}^t = 1$ and no unloading operation occurs ($ULD_{mnk}^t = 0$), the model must be reloaded to adapt its resource allocation. This condition applies to the following three scenarios:
\begin{itemize}
    \item Initial deployment: $R_{mnk}^{t-1} = 0$ to $R_{mnk}^{t} > 0$; 
    \item Resource increase:  $R_{mnk}^{t-1} > 0$ to $R_{mnk}^{t} > R_{mnk}^{t-1}$;
    \item Resource decrease: $R_{mnk}^{t-1} > 0$ to $0<R_{mnk}^{t} < R_{mnk}^{t-1}$.
\end{itemize}

In all such cases, the loading time overhead must be explicitly accounted for. 
To formally characterize this behavior, we introduce two binary state variables: $LD_{mnk}^t$ and $RLD_{mnk}^t$. The loading state variable $LD_{mnk}^t$ indicates whether the model was previously undeployed and is now being deployed, which can be defined as:
\begin{equation}
   LD_{mnk}^t = d_{mnk}^{t} \cdot (1 - d_{mnk}^{t-1}).
\end{equation}
The reloading state variable $RLD_{mnk}^t$ indicates whether the model remains deployed but undergoes a resource change, subject to the following constraints:
\begin{gather}
    RLD_{mnk}^t \leq RC_{mnk}^t, \\
    RLD_{mnk}^t \leq 1 - LD_{mnk}^t, \\
    RLD_{mnk}^t \leq 1 - ULD_{mnk}^t, \\
    RLD_{mnk}^t \geq RC_{mnk}^t + LD_{mnk}^t - ULD_{mnk}^t.
\end{gather}
The constraints ensure that the reloading state is only activated when a deployed model undergoes resource changes without being unloaded or newly deployed. Therefore, $\overline{TL}_{k}^{t}$ can be approximated as:
\begin{equation}
    \label{loadtime_tran2}
   \widetilde{TL}_{k}^{t} = \sum\limits_{m} (LD_{mnk}^{t} + R   LD_{mnk}^{t}) \cdot l_m.
\end{equation}

Finally, building on the derived convex latency function $\widetilde{L}_{mnk}^t$, static model generation quality $Q_{mn}$, model loaing time $\widetilde{TL}_{k}^{t}$ of GPU $k$, we reduce the original problem into the intra-node scheduling optimization at each edge node $n$:

\begin{gather}
\max \sum\limits_{m,k} p^t_{mnk}  Q_{mn}  \\
\text{s.t. } 
\begin{split}
\max\limits_{m,k} \big[ &\widetilde{L}_{mnk}^t(p_{mnk}^tB^t,R_{mnk}^t) + \widetilde{TL}_{k}^t \big] \leq L^t-TS_n^t,\forall t,
\end{split}
\label{eq:intra22}  \\
\sum\limits_{m} R_{mnk}^t \leq R_{k}, \forall n,k,t, \label{eq:intra33} \\
R_{mnk}^{t} \geq {d_{mnk}^t r_{m}}, \forall m,n,k,t, \label{eq:intra44}\\
R_{mnk}^{t} \leq d_{mnk}^t, \forall m,n,k,t. \label{eq:intra55} 
\end{gather}
Therefore, the optimization problem under consideration is a standard convex problem with linear constraints. In this way, it can be efficiently solved using existing convex optimization solvers, such as Gurobi or Mosek, within each edge node. 

\section{Evaluation}
\subsection{Evaluation Setups}
\label{5.1}

\textbf{Implementation Settings:} Our implementation employs VLLM 0.6.2 \cite{DBLP:conf/sosp/KwonLZ0ZY0ZS23}, a state-of-the-art LLM serving system optimized for high-throughput concurrent request processing, as the inference backend. For the policy network, the architecture features four fully-connected layers (256-128-64-action dim) with batch normalization and residual connections. For policy optimization, we configure PPO with a learning rate of $3e-4$ and a clipping threshold ($\epsilon$) of 0.02 to balance exploration and stability. For the feedback signal, the weight factor is set to be 1 and 0.5, respectively, determined through manual tuning to optimize overall system effectiveness. Retrieval operations leverage a Faiss-based vector database with a flat index for exact similarity search, retrieving the top-5 documents per query. To model heterogeneous edge environments, we deploy four nodes: two equipped with a single NVIDIA RTX 4090 GPU and two with dual NVIDIA RTX 4090 GPUs, simulating computational asymmetry. Note that our experiment targets micro-edge computing clusters, which offer more compute and memory than IoT devices yet remain resource-constrained relative to cloud infrastructure\footnote{Our architecture ensures hardware-agnostic operation: The hierarchical scheduler only requires model latency prediction and GPU memory constraints, decoupled from specific hardware. Additionally, the node profiler auto-adapts to devices ranging from high-performance GPUs to embedded systems (e.g., Jetson Orin), while remaining applicable to less powerful hardware through hardware-agnostic scheduling.}. Thus, RTX 4090, a cost-effective consumer-grade GPU, is a well-suited choice for real-world edge deployment \cite{DBLP:journals/corr/abs-2505-17052}.
Dynamic query patterns are emulated using real-world request traces from the ECW-New-App dataset \cite{HuangWZWZW23}, while domain-specific query bias is synthetically generated via Dirichlet sampling to simulate skewed per-slot query distributions typical of edge scenarios.

\textbf{Datasets and Query Synthesis:} To evaluate RAG performance across diverse domains in collaborative edge computing, we employ two datasets, each spanning six domains: \textbf{1)~DomainQA}: Six domains (biomedicine, finance, law, sports, technology, and travel) are selected from the BAAI industry corpus series\footnote{https://huggingface.co/BAAI}. Domain-specific corpora at the edge are constructed by sampling representative data from each domain. Using DeepSeek-V3 API, we automate the generation of 3,000 question-answer pairs per domain, ensuring alignment with semantic and contextual information of the extracted corpora. \textbf{2)~Personalized-Proactive-Conversations (PPC)\footnote{https://huggingface.co/datasets/erbacher/personalized-proactive-conversations/viewer}}: User conversations with a cloud-based LLM assistant are filtered to retain six distinct identity profiles (student, teacher, parent, engineer, chef, and writer), reflecting diverse practical scenarios. Through in-depth dialogue analysis, user inquiries are transformed into a series of independent query statements, preserving the authenticity of needs for reusing historical conversations in real-world applications.

\textbf{Edge-data Partition:} To efficiently characterize the heterogeneous data distribution at the edge, similar to \cite{KarimireddyKMRS20}, the edge-data partition follows a dual-distribution paradigm: s\% of the data samples from each client are distributed as i.i.d. document chunks spanning all categories, while the remaining (100 - s)\% are allocated in a non-i.i.d. fashion where each node is assigned data from three predefined domains. To explicitly model cross-node knowledge-sharing scenarios, we introduce an overlapping factor that scales both i.i.d. and non-i.i.d. data portions, thereby facilitating knowledge-sharing potential through controlled dataset intersections between nodes.

\textbf{Edge LLMs:}
We construct a heterogeneous model pool comprising multiple parameter-efficient variants of widely adopted architectures. Specifically, we curate three prominent open-source model series optimized for edge deployment: LLaMA \cite{grattafiori2024llama}, Qwen \cite{qwen2025qwen25technicalreport}, and Falcon \cite{almazrouei2023falcon}, spanning parameter sizes of 1B/1.5B, 3B, and 7/8B. This multi-scale parameterization across architectures underscores the diverse deployment requirements inherent to edge computing systems.

\textbf{Evaluation Metrics:}
To evaluate generation quality, we employ common lexical and semantic metrics, including Rouge \cite{lin2004rouge}, Bleu \cite{papineni2002bleu}, Meteor \cite{banerjee2005meteor}, BERTScore \cite{zhang2019bertscore} to
quantify the similarity between LLM responses and the references, where higher scores indicate higher quality. In terms of the query completion performance, we measure the DropRate, i.e., the fraction of queries violating latency constraints per time slot.

\subsection{Evaluation Results and Analysis}




\begin{table*}[htb]
\centering
\small 
\caption{Performance comparison of different query allocation method.}
\begin{tabular}{|l|l|c|c|c|c|c|c|}
\hline
\multicolumn{2}{|c|}{} & \textbf{ROUGE-1} & \textbf{ROUGE-2} & \textbf{ROUGE-L} & \textbf{BLEU-4} & \textbf{METEOR} & \textbf{BERTScore} \\
\hline
\multirow{5}{*}{\textbf{DomainQA}} 
& \textbf{Random}  & 0.477 & 0.322 & 0.438 & 0.249 & 0.460 & 0.698 \\
& \textbf{MAB}  & 0.569 & 0.421 & 0.531 & 0.337 & 0.560 & 0.756  \\
& \textbf{PPO} & 0.626 & 0.486 & 0.589 & 0.398 & 0.621 & 0.788 \\
& \textbf{Oracle}  & 0.648 & 0.506 & 0.609 & 0.413 & 0.650 & 0.806 \\
\hline
\multirow{5}{*}{\textbf{PPC}} 
& \textbf{Random}  & 0.403 & 0.211 & 0.373 & 0.151 & 0.355 & 0.682 \\
& \textbf{MAB}  & 0.499 & 0.316 & 0.471 & 0.238 & 0.456 & 0.738 \\
& \textbf{PPO}  & 0.555 & 0.377 & 0.528 & 0.289 & 0.516 & 0.767 \\
& \textbf{Oracle}  & 0.569 & 0.388 & 0.541 & 0.296 & 0.526 &  0.777 \\

\hline
\end{tabular}
\label{tab:query-match}
\end{table*}

\textbf{Query-edge Matching Impacts Generation Quality.}
To verify the effectiveness of our
online query identification method, we benchmark against three baselines: \textbf{1)~Random Allocation} randomly routes queries to edge nodes without semantic awareness. \textbf{2)~MAB-based Allocation} adopts the LinUCB algorithm \cite{li2010contextual} to allocate queries based on historical performance and uncertainty estimates,  
rather than the feature extraction of a neural network. \textbf{3)~Oracle Allocation} assigns user queries to the optimal edge node with perfect knowledge of corpus distributions. 
Table~\ref{tab:query-match} compares the utility of our method against baselines on the DomainQA and PPC datasets. The results demonstrate that our approach significantly outperforms both Random and MAB-based allocation strategies across all metrics (around 4.23\%--59.84\% improvements on DomainQA dataset and 3.93\%--91.39\% improvements on PPC dataset). Notably, it achieves performance comparable to the Oracle baseline, highlighting its robustness in practical, privacy-sensitive edge environments. This superiority stems from our method’s capacity to model high-dimensional query semantics and dynamically align them with distributed corpus data via iterative updates in dynamic edge environments. In contrast, random allocation --- agnostic to data semantics --- induces query-node mismatches, precluding LLMs from exploiting high-quality relevant retrieval data. Besides, MAB-based allocation fails to model high-dimensional query features and struggles to adapt to environmental dynamics, resulting in suboptimal performance.

\begin{figure*}[ht]
  \centering
  \begin{subfigure}{0.24\textwidth} 
    \includegraphics[width=\linewidth, height = 0.75\linewidth]{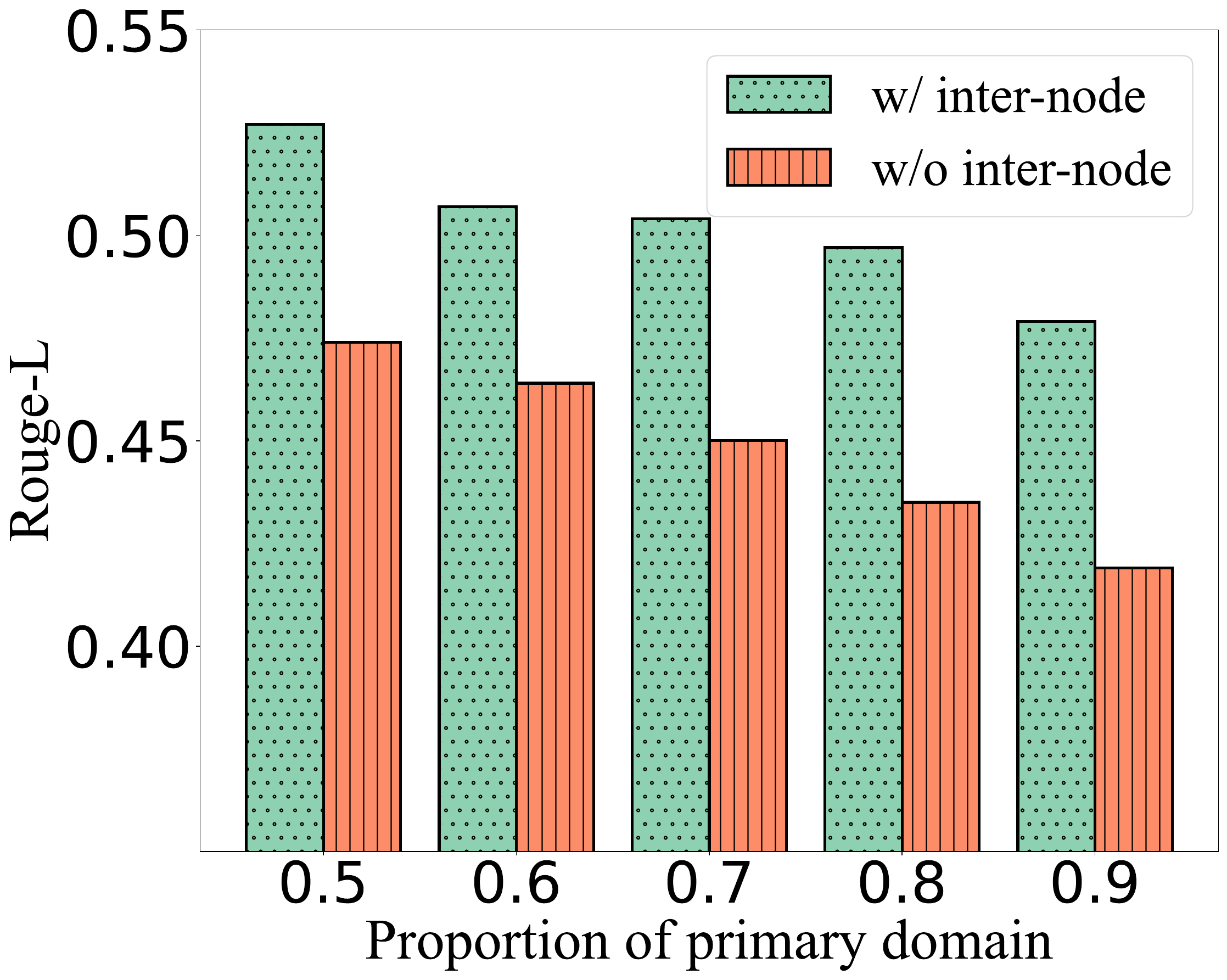}
    \caption{Rouge-L performance on DomainQA dataset.}
    \label{fig:inter_node1}
  \end{subfigure}
  \hfill
  \begin{subfigure}{0.24\textwidth}
    \includegraphics[width=\linewidth, height = 0.75\linewidth]{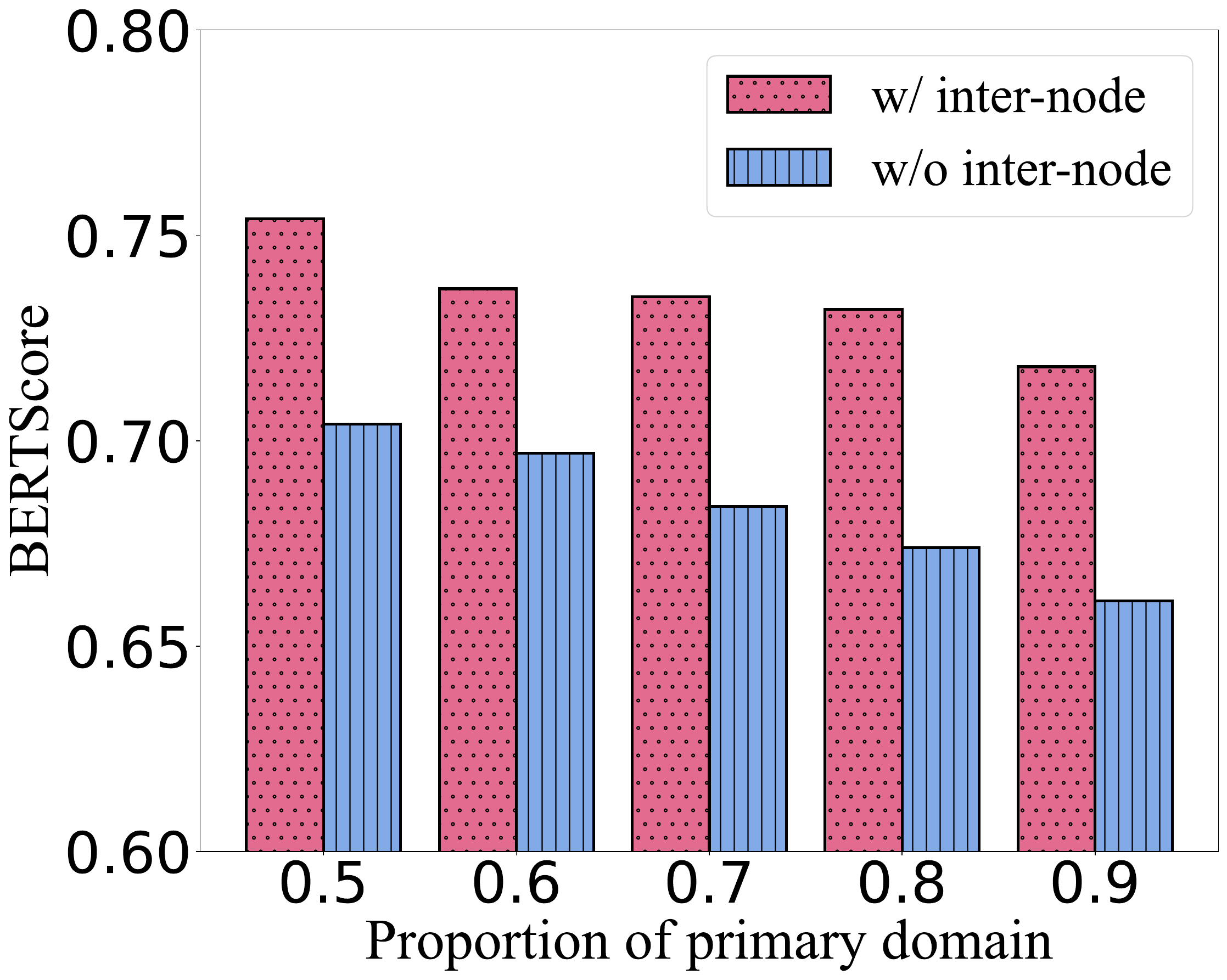}
    \caption{BertScore performance on DomainQA dataset.}
    \label{fig:inter_node2}
  \end{subfigure}
  \hfill
  \begin{subfigure}{0.24\textwidth}
    \includegraphics[width=\linewidth, height = 0.75\linewidth]{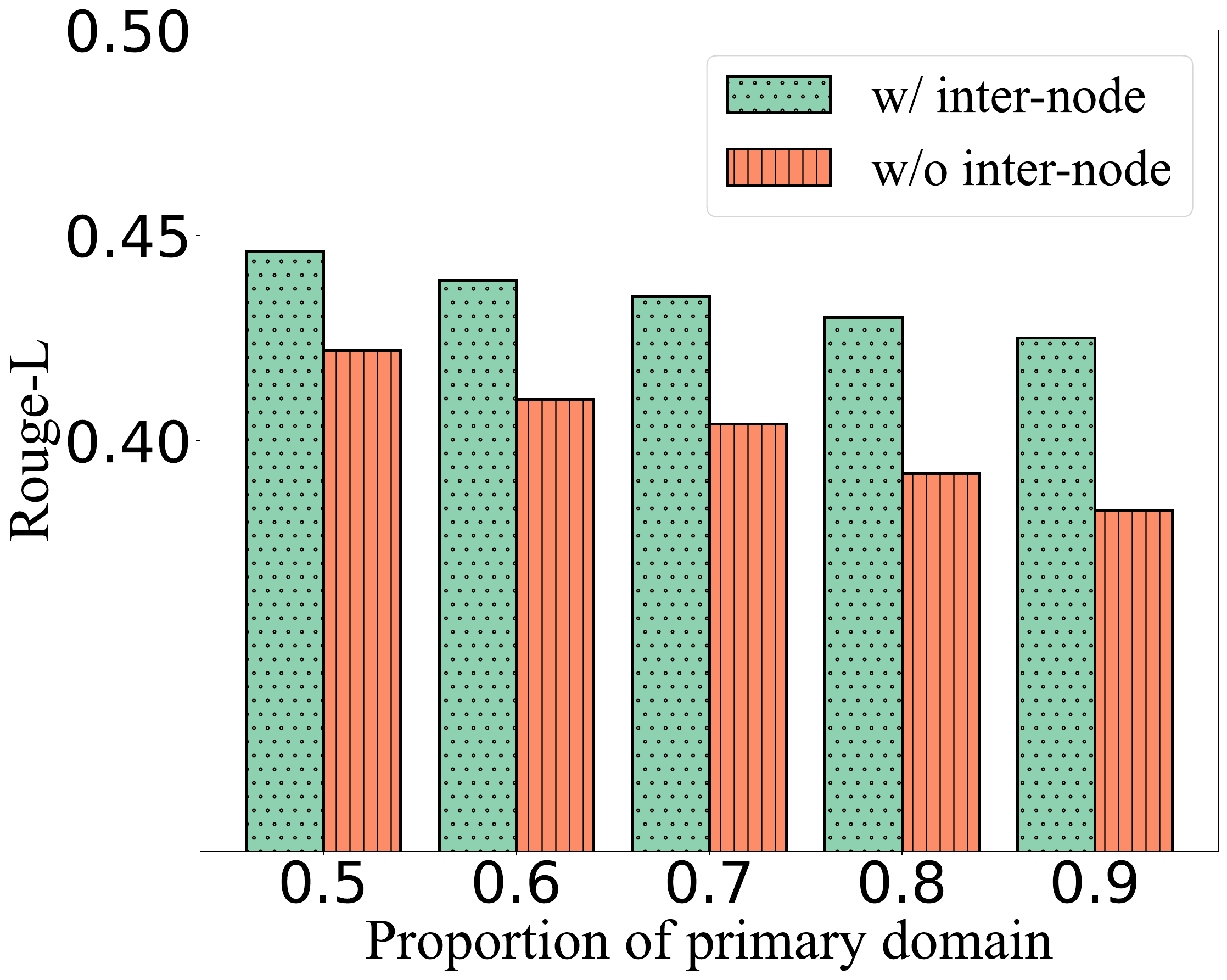}
    \caption{Rouge-L performance on PPC dataset.}
    \label{fig:inter_node3}
  \end{subfigure}
  \hfill
  \begin{subfigure}{0.24\textwidth}
    \includegraphics[width=\linewidth, height = 0.75\linewidth]{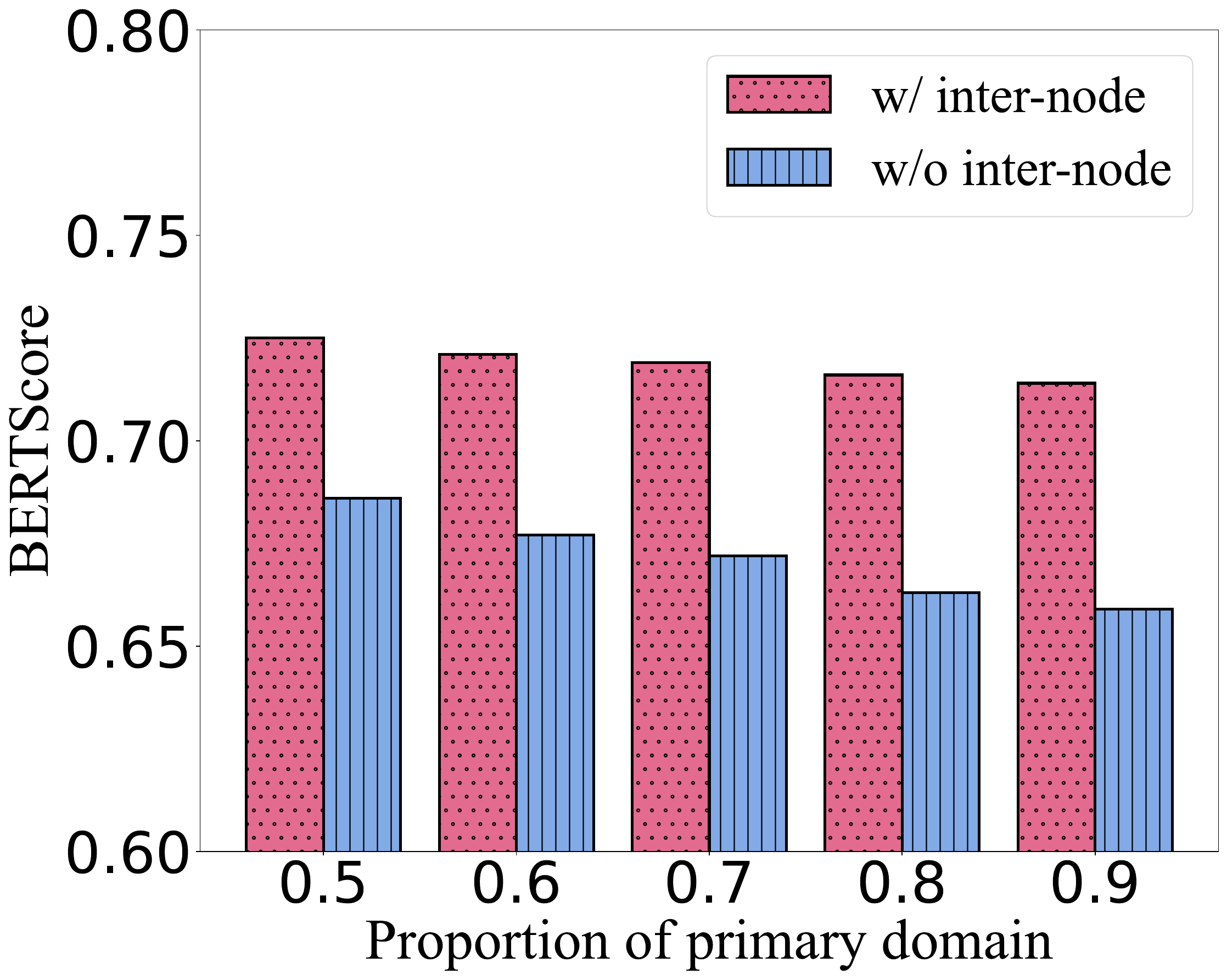}
    \caption{BertScore performance on PPC dataset.}
    \label{fig:inter_node4}
  \end{subfigure}
  \caption{Generation quality of different scheduling strategies.}
  \label{fig:inter_node_all}
  \vspace{-10pt}
\end{figure*}

\textbf{Workload Balancing Impacts on LLM Performance.}
Under controlled experimental conditions with fixed query loads and stringent latency constraints (DomainQA: 2000 queries, \SI{15}{s} latency; PPC: 1500 queries, \SI{15}{s} latency), we systematically evaluated performance by designating one domain as primary and progressively increasing its representation within query batches to simulate workload skew.

For the DomainQA dataset, inter-node method consistently outperformed w/o inter-node approaches, demonstrating mean advantages of 12.65\% in Rouge-L and 7.71\% in BertScore across varying domain concentrations.  When the primary domain proportion increased from 0.5 to 0.9, the Rouge-L score decreased from 0.527 to 0.485 with inter-node scheduling, compared to a decrease from 0.474 to 0.416 without inter-node coordination. Similarly, the BertScore decreased from 0.754 to 0.718 with inter-node scheduling versus a decrease from 0.704 to 0.661 without.
Similar trends emerged in PPC evaluations, where inter-node scheduling maintained 8.21\% and 7.13\% average leads in Rouge-L and BertScore, respectively. With inter-node scheduling, the Rouge-L score decreased from 0.446 to 0.425, while without inter-node coordination it decreased from 0.422 to 0.383. The BertScore showed greater stability, decreasing from 0.725 to 0.714 with inter-node scheduling compared to a decrease from 0.686 to 0.659 without.

The findings highlight two key observations: 
(i) Performance degrades with increasing domain skew across all configurations, underscoring the fundamental difficulties of biased query distributions. 
(ii) The dynamic exploration in inter-node scheduling alleviates this degradation through cross-node optimization. 
The architecture demonstrates robust real-world applicability by maintaining sub-threshold latency while ensuring stable performance and efficient data utilization --- particularly valuable in environments with natural query skewness. This resilience stems from the system's probabilistic balancing mechanism between immediate scheduling requirements and exploratory node sampling, which prevents local optima that worsen performance under skewed workloads.

\begin{table*}[htb]
\centering
\small 
\caption{Performance (R-1: ROUGE-1, R-2: ROUGE-2, R-L: ROUGE-L) comparison of different allocation methods within node over varying $L$ (s) values.  \textcolor{CellFst}{\rule{3mm}{3mm}} indicate the best performance, and \textcolor{CellSnd}{\rule{3mm}{3mm}} indicate the second-best performance for each metric.} 
\begin{tabular}{@{}|l|c|c|c|c|c|c|c|c|c@{}|}
\hline
 \multicolumn{3}{|c|}{}  & \textbf{R-1} & \textbf{R-2} & \textbf{R-L} & \textbf{BLEU-4} & \textbf{METEOR} & \textbf{BERTScore} & \textbf{DropRate($\%$)} \\
\hline
\multirow{15}{*}{\textbf{DomainQA}} 
& \multirow{5}{*}{\textbf{L=5}} & \textbf{Small-Param} & \cellcolor{CellSnd}\textbf{0.552} & \cellcolor{CellSnd}\textbf{0.408} & \cellcolor{CellSnd}\textbf{0.520} & \cellcolor{CellSnd}\textbf{0.323} & \cellcolor{CellSnd}\textbf{0.552} & \cellcolor{CellSnd}\textbf{0.743} & 3.57 \\
& & \textbf{Mid-Param} & 0.204 & 0.159 & 0.191 & 0.132 & 0.201 & 0.459 & 66.71 \\
& & \textbf{Mixed-Param.1}  & 0.394 & 0.299 & 0.371 & 0.242 & 0.392 & 0.611 & 34.44 \\
& & \textbf{Mixed-Param.2} & 0.214 & 0.168 & 0.203 & 0.138 & 0.214 & 0.468 & 65.72 \\
& & \textbf{Intra-node} & \cellcolor{CellFst}\textbf{0.555} & \cellcolor{CellFst}\textbf{0.410} & \cellcolor{CellFst}\textbf{0.523} & \cellcolor{CellFst}\textbf{0.324} & \cellcolor{CellFst}\textbf{0.556} & \cellcolor{CellFst}\textbf{0.746} & 2.81 \\
\cline{2-10}
& \multirow{5}{*}{\textbf{L=10}} & \textbf{Small-Param} & 0.585 & 0.435 & 0.552 & 0.345 & 0.588 & 0.768 & 0.02 \\
& & \textbf{Mid-Param} & 0.588 & 0.439 & 0.547 & 0.355 & 0.584 & 0.769 & 0.01 \\
& & \textbf{Mixed-Param.1}  & 0.605 & 0.459& 0.569 & 0.370 & 0.606 & 0.779 & 0.02 \\
& & \textbf{Mixed-Param.2} & \cellcolor{CellSnd}\textbf{0.611} & \cellcolor{CellSnd}\textbf{0.466} & \cellcolor{CellSnd}\textbf{0.574} & \cellcolor{CellSnd}\textbf{0.377} & \cellcolor{CellSnd}\textbf{0.612} & \cellcolor{CellSnd}\textbf{0.782} & 0.04 \\
& & \textbf{Intra-node} & \cellcolor{CellFst}\textbf{0.626} & \cellcolor{CellFst}\textbf{0.484} & \cellcolor{CellFst}\textbf{0.587} & \cellcolor{CellFst}\textbf{0.397} & \cellcolor{CellFst}\textbf{0.624} & \cellcolor{CellFst}\textbf{0.789} & 0.14 \\
\cline{2-10}
& \multirow{5}{*}{\textbf{L=15}} & \textbf{Small-Param} & 0.585 & 0.435 & 0.552 & 0.345 & 0.588 & 0.768 & 0.00 \\
& & \textbf{Mid-Param} & \cellcolor{CellSnd}\textbf{0.626} & \cellcolor{CellSnd}\textbf{0.483} & \cellcolor{CellSnd}\textbf{0.586} & \cellcolor{CellSnd}\textbf{0.395} & 
\cellcolor{CellSnd}\textbf{0.623} & \cellcolor{CellSnd}\textbf{0.790} & 0.00 \\
& & \textbf{Mixed-Param.1} & 0.606 & 0.459 &  0.569 & 0.370 & 0.606 & 0.779 & 0.00 \\
& &\textbf{Mixed-Param.2} & 0.611 & 0.466 & 0.574 & 0.377 & 0.612 & 0.782 & 0.00 \\
& & \textbf{Intra-node} & \cellcolor{CellFst}\textbf{0.637} & \cellcolor{CellFst}\textbf{0.499} & \cellcolor{CellFst}\textbf{0.598} & \cellcolor{CellFst}\textbf{0.412} & \cellcolor{CellFst}\textbf{0.635} & \cellcolor{CellFst}\textbf{0.796} & 0.00 \\
\hline
\multirow{15}{*}{\textbf{PPC}} 
& \multirow{5}{*}{\textbf{L=5}} & \textbf{Small-Param} & \cellcolor{CellFst}\textbf{0.478} & \cellcolor{CellSnd}\textbf{0.296} & \cellcolor{CellFst}\textbf{0.453} & \cellcolor{CellFst}\textbf{0.217} & \cellcolor{CellSnd}\textbf{0.442} & \cellcolor{CellFst}\textbf{0.722} & 1.84 \\
& & \textbf{Mid-Param} & 0.310 & 0.215 & 0.295 & 0.163 & 0.278 & 0.557 & 44.34 \\
& & \textbf{Mixed-Param.1}  & 0.396 & 0.259 & 0.376 & 0.192 & 0.359 & 0.640 & 23.28 \\
& & \textbf{Mixed-Param.2} & 0.214 & 0.145 & 0.204 & 0.109 & 0.197 & 0.476& 60.78 \\
& & \textbf{Intra-node } & \cellcolor{CellSnd}\textbf{0.478} & \cellcolor{CellFst}\textbf{0.297} & \cellcolor{CellSnd}\textbf{0.453} & \cellcolor{CellSnd}\textbf{0.216}  & \cellcolor{CellFst}\textbf{0.444} & \cellcolor{CellSnd}\textbf{0.722} & 1.96 \\
\cline{2-10}
& \multirow{5}{*}{\textbf{L=10}} & \textbf{Small-Param} & 0.507 & 0.324 & 0.481 & 0.238 & 0.473 & 0.741 & 0.01 \\
& & \textbf{Mid-Param} & 0.517 & 0.339 & 0.490 & 0.254 & 0.464 & 0.747 & 0.02 \\
& & \textbf{Mixed-Param.1} & 0.529 & 0.352 & 0.504 & 0.264 & 0.489 & 0.752 & 0.71 \\
& & \textbf{Mixed-Param.2} & \cellcolor{CellSnd}\textbf{0.536} & \cellcolor{CellSnd}\textbf{0.359} & \cellcolor{CellSnd}\textbf{0.510} & \cellcolor{CellSnd}\textbf{0.272} & \cellcolor{CellSnd}\textbf{0.500} & \cellcolor{CellSnd}\textbf{0.756} & 0.63 \\
& & \textbf{Intra-node} & \cellcolor{CellFst}\textbf{0.550} & \cellcolor{CellFst}\textbf{0.374} & \cellcolor{CellFst}\textbf{0.524} & \cellcolor{CellFst}\textbf{0.287} & \cellcolor{CellFst}\textbf{0.513} & \cellcolor{CellFst}\textbf{0.764} & 0.46 \\
\cline{2-10}
& \multirow{5}{*}{\textbf{L=15}} & \textbf{Small-Param} & 0.509 & 0.325 & 0.484 & 0.239 & 0.474 & 0.742 & 0.00 \\
& & \textbf{Mid-Param} & \cellcolor{CellSnd}\textbf{0.554} & \cellcolor{CellSnd}\textbf{0.380} & \cellcolor{CellSnd}\textbf{0.527} & \cellcolor{CellSnd}\textbf{0.290} & \cellcolor{CellSnd}\textbf{0.506} & \cellcolor{CellSnd}\textbf{0.767} & 0.00 \\
& & \textbf{Mixed-Param.1}  & 0.533 & 0.355 & 0.507 & 0.266 & 0.492 & 0.755 & 0.00 \\
& & \textbf{Mixed-Param.2} & 0.540 & 0.361 & 0.514 & 0.273 & 0.503 & 0.759 & 0.00 \\
& & \textbf{Intra-node} & \cellcolor{CellFst}\textbf{0.564} & \cellcolor{CellFst}\textbf{0.388} & \cellcolor{CellFst}\textbf{0.537} & \cellcolor{CellFst}\textbf{0.301} & \cellcolor{CellFst}\textbf{0.529} & \cellcolor{CellFst}\textbf{0.773} & 0.00 \\
\hline
\end{tabular}
\label{tab:intra_performance}
\end{table*}

\begin{figure*}[ht]
  \centering
  \begin{subfigure}{0.24\textwidth} 
    \includegraphics[width=\linewidth, height = 0.75\linewidth]{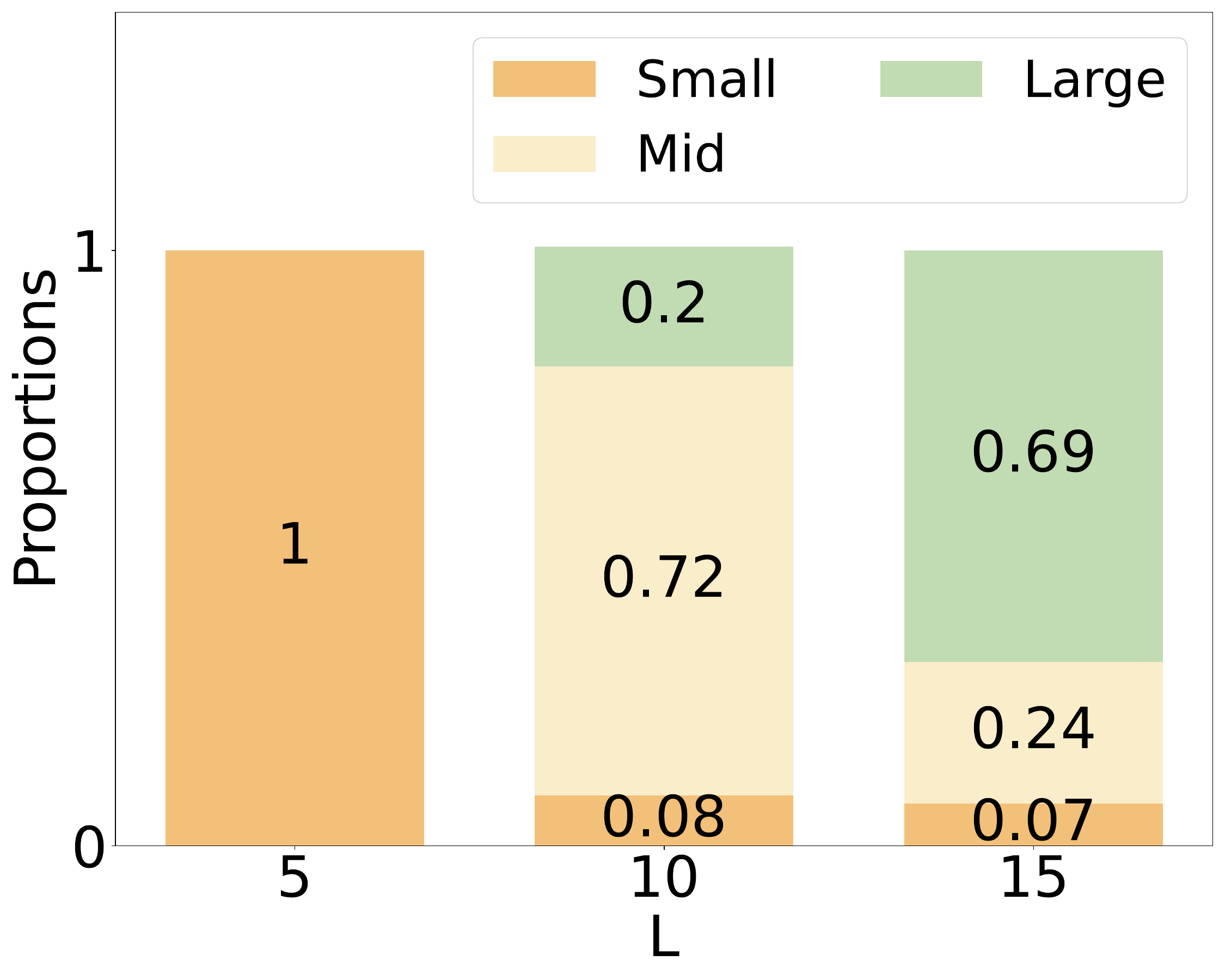}
    \caption{Proportion of queries on DomainQA dataset.}
\label{fig:query_proporation_domainqa}
  \end{subfigure}
  \hfill
  \begin{subfigure}{0.24\textwidth}
    \includegraphics[width=\linewidth, height = 0.75\linewidth]{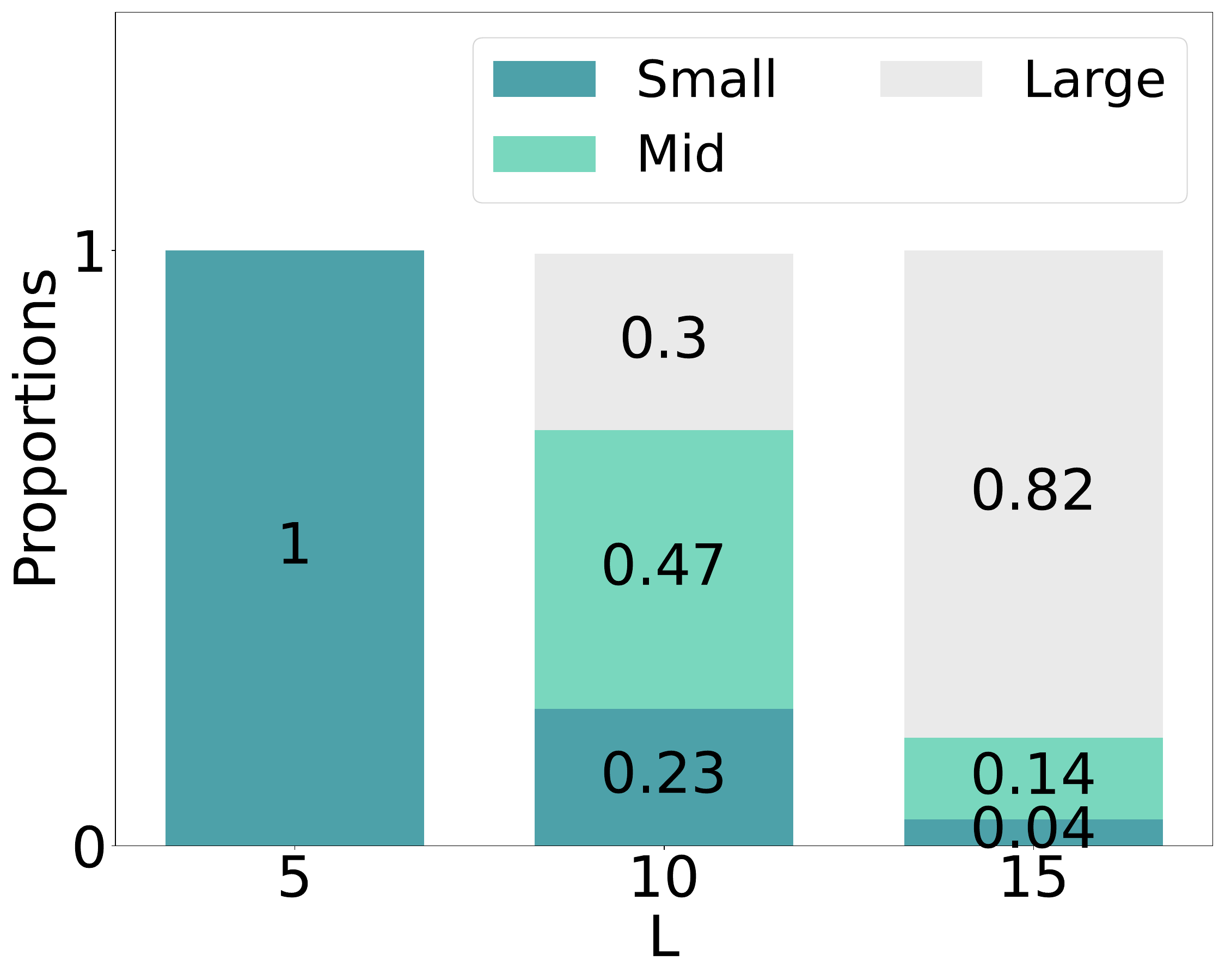}
    \caption{Proportion of resources on DomainQA dataset.}
    \label{fig:resource_proportion_domainqa}
  \end{subfigure}
  \hfill
  \begin{subfigure}{0.24\textwidth}
    \includegraphics[width=\linewidth, height = 0.75\linewidth]{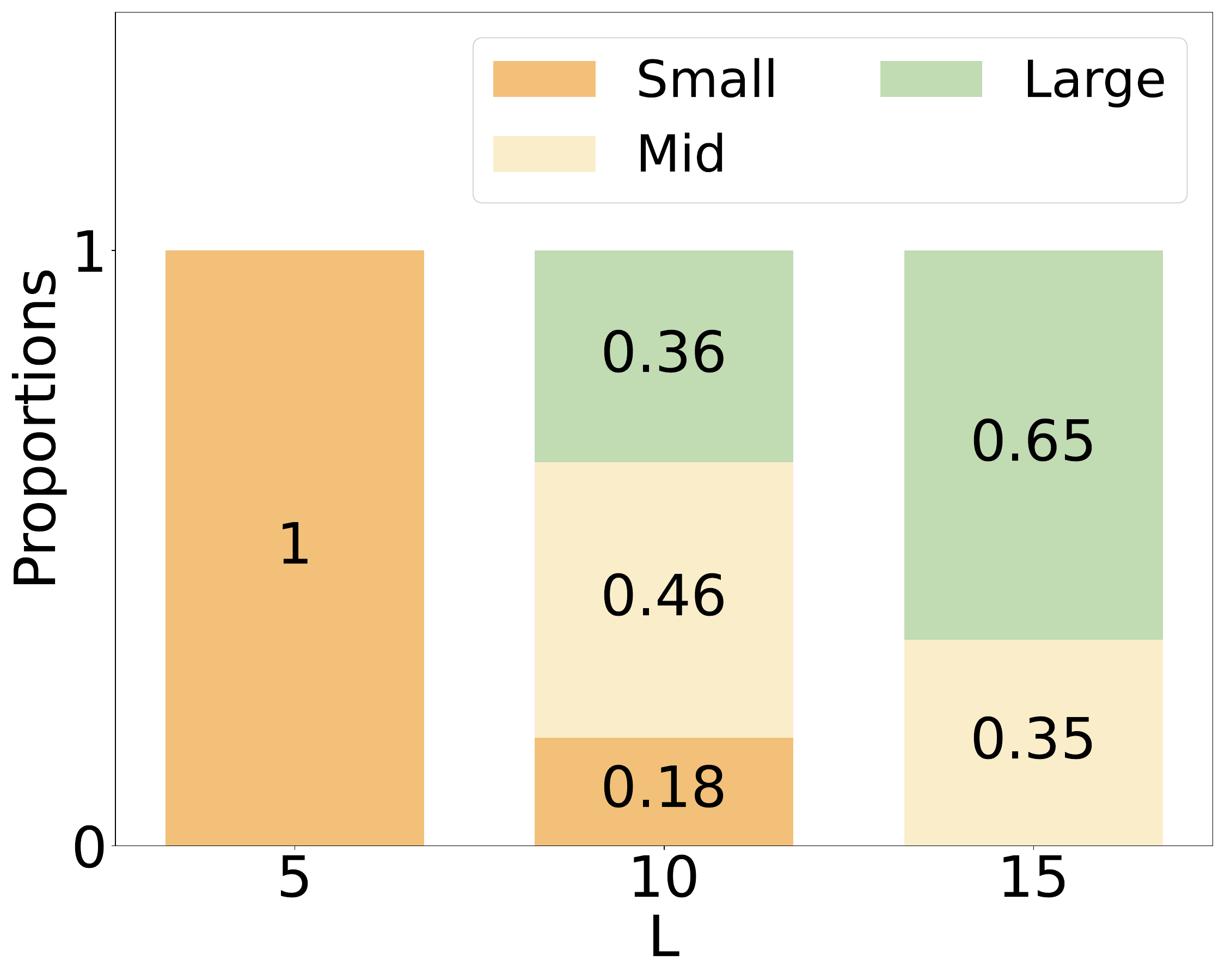}
    \caption{Proportion of queries on PPC dataset.}
    \label{fig:query_proporation_ppc}
  \end{subfigure}
  \hfill
  \begin{subfigure}{0.24\textwidth}
    \includegraphics[width=\linewidth, height = 0.75\linewidth]{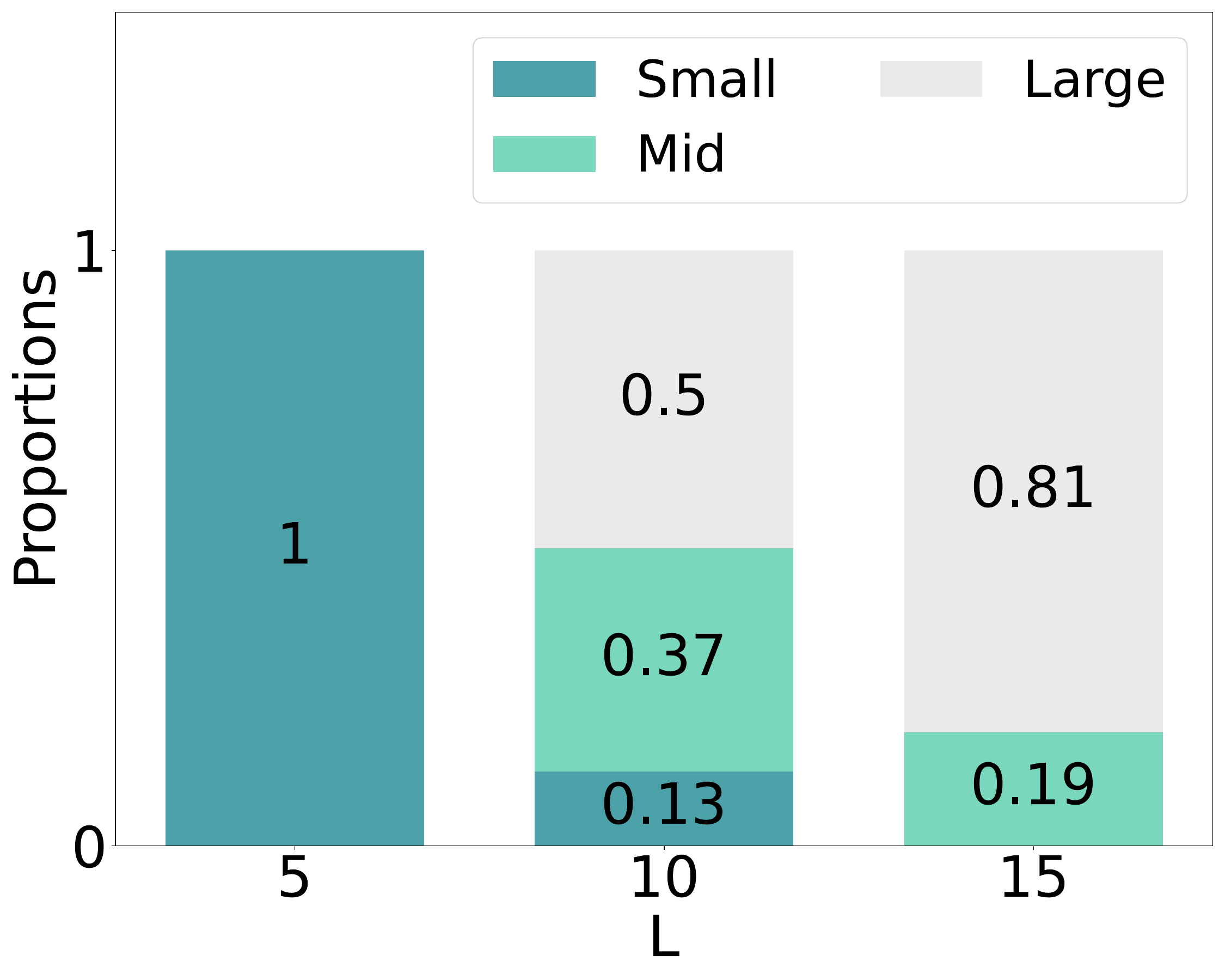}
    \caption{Proportion of resources on PPC dataset.}
    \label{fig:resource_proportion_ppc}
  \end{subfigure}
  \caption{Proportion with different model sizes.}
  \label{fig:proportion}
  \vspace{-10pt}
\end{figure*}

\textbf{Robustness in Different Latency SLOs.} 
We introduced baselines with different model combinations  to serve user queries: \textbf{1)~Small-Param}: each node deploys only small-parameter models (1B/1.5B); \textbf{2)~Mid-Param}: each node deploys only medium-parameter models (3B); \textbf{3)~Mixed-Param.1}: each GPU deploys both small and medium-parameter models with fixed query allocation proportion and resource propotion; \textbf{4)~Mixed-Param.2}: single-GPU nodes deploy small and medium-parameter models, while dual-GPU nodes allocate one GPU to small/medium models and the other to large-parameter models (7B/8B). Among these,  queries are evenly distributed among deployed models.

Experiments were conducted on the DomainQA (500 queries) and PPC (400 queries) datasets under latency constraints of \SI{5}{s}, \SI{10}{s}, and \SI{15}{s}, translating to per-query processing budgets of 10--30~\si{ms} and 12.5--37.5~\si{ms}, respectively.  This setup allowed us to assess RAG system performance under progressively relaxed latency requirements. As shown in Table~\ref{tab:intra_performance}, intra-node scheduling demonstrates exceptional adaptability across all latency conditions.

\textbf{Under strict latency constraints ($L=$ \SI{5}{s})}, our method maintains a drop rate below 3\%, with all metrics ranking among the top two across both datasets. The Small-Param configuration, exclusively equipped with small models, performs particularly well, as its design inherently suits latency-sensitive tasks. In contrast, Mid-Param and Mixed-Param.2 exhibit catastrophic drop rates (44.34\%--66.71\%) due to their inability to dynamically allocate queries to smaller models or reallocate resources. This results in severe degradation in LLM generation quality. Meanwhile, Mixed-Param.1, although incorporating smaller models to share the processing burden, still exhibits unacceptably poor performance (23.28\%--34.44\%).

\textbf{Under moderate latency constraints ($L=$ \SI{10}{s})}, our method achieves superior performance across all metrics with a near-zero drop rate. 
The relaxed latency constraints enable the intra-node scheduling mechanism to strategically allocate more queries to larger models, enhancing overall system performance. 
Meanwhile, the Mixed-Param.2 configuration achieves notable improvements by co-deploying small-, medium-, and large-parameter models, which enables an effective balance between latency and generation quality.
By assigning a portion of queries to smaller models to avoid severe timeouts, and reserving larger models for acceptable LLM responses, the system ensures both timely responses and satisfactory generation quality. In contrast, Mixed-Param.1 only deploys small and medium-parameter models, resulting in relatively lower generation quality.
Notably, Small-Param's performance plateaus, as its exclusive reliance on small models prevents it from capitalizing on the relaxed latency constraints. This limitation highlights the critical role of model diversity in adaptive scheduling frameworks.
For Mid-Param, while the inter-node scheduling's capacity exploration mechanism effectively mitigates query overload, the exclusive use of medium-parameter models eliminates small-model fallback capabilities, preventing optimal node assignment and consequently compromising overall system performance.
The empirical evaluation reveals a fundamental trade-off: while an over-reliance on smaller models ensures higher reliability under stringent latency constraints, it inherently restricts the achievable quality improvements when requirements are loosened. These findings provide compelling evidence for adopting dynamic resource allocation mechanisms, which enables adaptive performance optimization across varying operational conditions, thereby balancing efficiency and effectiveness in real-time systems.

\textbf{Under relaxed latency constraints ($L=$ \SI{15}{s})},
our method continues to demonstrate consistent superiority, maintaining exceptional service quality while simultaneously maximizing generation fidelity through intelligent adaptive model deployment. Notably, Mid-Param benefits from the further relaxed latency constraints, eliminating the need for inter-node scheduling with sub-optimal results, which leads to a significant improvement in generation quality. In contrast, other strategies show minimal to no performance gains. 


The comprehensive experimental evaluation conclusively demonstrates that our intra-node scheduling paradigm consistently outperforms conventional approaches across the entire latency constraint spectrum. Unlike these baselines, which either degrade significantly under stringent temporal requirements or fail to capitalize on relaxed constraints to enhance output quality, our solution maintains exceptional stability while simultaneously preserving high generation fidelity. These results not only validate the proposed architecture's robustness but also establish a new benchmark for adaptive resource management in distributed inference systems.

\textbf{Dynamic Adjustability Across Different Latency SLOs.} To analyze the adaptive behavior of our intra-node scheduling strategy across varying latency conditions, we systematically tracked both query and resource allocation patterns across different model sizes (Fig.~\ref{fig:proportion}). The results reveal three distinct operational regimes:  
\textbf{Under strict latency constraints}, the scheduler demonstrates intelligent prioritization by directing all of queries to small-parameter models, thereby guaranteeing reliable adherence to stringent Service Level Objectives (SLOs).
\textbf{Under moderate latency constraints},
the system automatically rebalances toward quality optimization:  As shown in Fig.~\ref{fig:query_proporation_domainqa} and \ref{fig:query_proporation_ppc}, medium-parameter models handle 72\% and 46\% of queries on DomainQA and PPC, respectively, consuming a proportional share of resources (47\%, 37\%). Concurrently, large-parameter models process 28\% (30\% resources) and 54\% (50\% resources) of the workloads.
This graded reallocation underscores its adaptive nature, leveraging increased latency headroom for higher-quality outputs. Notably, resource allocation patterns exhibit non-linear scaling--large-parameter models receive disproportionately higher resources per query, reflecting their greater computational demands.
\textbf{Under relaxed latency constraints}, the scheduling strategy fully exploits the available latency budget, directing the majority of queries (65\% and 69\%) to large-parameter models to maximize response quality.
The observed dynamics demonstrate our scheduler's capability to make principled latency-quality trade-offs. Particularly noteworthy is the consistent correlation between query distribution and resource allocation across both datasets, validating the robustness of our adaptive mechanisms.


\section{Related Work}
\subsection{RAG-based LLM}
Significant efforts have been made to optimize RAG systems
from various perspectives. For instance, self-RAG \cite{asai2023self} introduces a reflection token and self-reflection mechanism, enabling LLMs to dynamically retrieve, generate, and evaluate content during the generation process. BlendedRAG \cite{sawarkar2024blended} enhances system accuracy by integrating dense vector indexing from semantic search with sparse encoder indexing, utilizing a hybrid query strategy. Beyond improving retrieval quality, another major thrust of research focuses on enhancing the execution efficiency of RAG systems. EdgeRAG \cite{seemakhupt2024edgerag} addresses performance issues on edge devices by pruning embedding vectors and employing caching strategies to reduce retrieval latency and optimize memory usage. PipeRAG \cite{jiang2024piperag} improves generation efficiency through co-designing algorithms and systems, enabling pipeline parallelism and dynamic retrieval adjustments. RAGCache \cite{DBLP:journals/corr/abs-2404-12457} significantly reduces latency and boosts throughput by efficiently caching and reusing intermediate states of retrieved documents through a multi-level dynamic caching mechanism. However, these systems primarily target single-node optimization. In contrast, our CoEdge-RAG focuses on multi-node architectures, enhancing the coordination efficiency of distributed RAG systems through hierarchical scheduling.
\subsection{Collaborative Edge LLM Serving}
PICE \cite{zhan2025pice} introduces a semantic-driven progressive inference system for edge-cloud environments, employing dynamic scheduling strategies to achieve high throughput and low-latency parallel edge inference. Ding et al. \cite{ding2024enhancing} enhance on-device LLM inference by constructing a local external datastore from historical cloud-based LLM interactions, leveraging k-nearest neighbor language modeling method. Helix \cite{mei2024helix} optimizes LLM serving in heterogeneous GPU clusters through layer placement and request scheduling formulated as a Max-Flow problem, ensuring high throughput and low latency. Edgeshard \cite{zhang2024edgeshard} partitions LLMs into shards and utilizes dynamic programming to optimize edge deployment efficiency. These efforts primarily focus on computational or architectural optimizations. Crucially, the rich, distributed data at the edge --- a pivotal resource for refining LLM capabilities --- remains underutilized in existing methodologies.

\section{Conclusion}
This paper presents CoEdge-RAG, a hierarchical framework for optimizing retrieval-augmented LLMs in collaborative edge environments. By integrating an online query identification mechanism, dynamic inter-node workload balancing, and intra-node resource allocation, CoEdge-RAG efficiently unleashes the potential of distributed data and heterogeneous computational resources to enhance edge-based LLM inference. Rigorous evaluations demonstrate the efficacy and robustness of CoEdge-RAG in handling dynamic and diverse queries, paving the way for its widespread adoption in future ubiquitous edge intelligence applications.

\

\bibliographystyle{IEEEtran} 
\bibliography{references} 



\end{document}